\documentclass[useAMS,usenatbib]{mn2e}
\usepackage{threeparttable}
\usepackage{color}
\usepackage[]{txfonts}
\def\simless{\mathbin{\lower 3pt\hbox
{$\rlap{\raise 5pt\hbox{$\char'074$}}\mathchar"7218$}}}   %< or of order
\def\simmore{\mathbin{\lower 3pt\hbox
{$\rlap{\raise 5pt\hbox{$\char'076$}}\mathchar"7218$}}}   %> or of order
\newcommand{\eqb}{\begin{eqnarray}}
\newcommand{\eqe}{\end{eqnarray}}
\newcommand{\mel}{m_{\rm e}}
\newcommand{\mpr}{m_{\rm p}}

\newcommand{\pg}{ {\rm p}\pi}

\newcommand{\rb}{R}

\newcommand{\sth}{\sigma_{\rm T}}
\newcommand{\lpinj}{\ell_{\rm p, inj}}
\newcommand{\leinj}{\ell_{\rm e, inj}}

\newcommand{\gpmx}{\gamma_{\rm p,max}}

\newcommand{\tpesc}{t_{\rm p,esc}}
\newcommand{\teesc}{t_{\rm e,esc}}

\newcommand{\rxs}{1~RXS~J054357.3-553206}
\newcommand{\ygn}{Y_{\nu \gamma}}
\usepackage{graphicx}
\usepackage{epsfig} 
\usepackage{epstopdf}

\title[Testing the BL~Lac-IceCube neutrino connection]
{Photohadronic origin of $\gamma$-ray BL~Lac emission: implications for IceCube neutrinos}

\author[]
{M. Petropoulou$^{1}$\thanks{E-mail: mpetropo@purdue.edu}\thanks{Einstein Postdoctoral Fellow}, S.~Dimitrakoudis$^{2}$, P. Padovani$^3$, A. Mastichiadis$^4$, E. Resconi$^5$
\\
$^{1}$Department of Physics and Astronomy, Purdue University, 525 Northwestern
Avenue, West Lafayette, IN 47907, USA\\
$^2$Institute for Astronomy, Astrophysics, Space Applications \& Remote Sensing, National Observatory of Athens, 15 236 Penteli, Greece \\
$^3$European Southern Observatory, Karl-Schwarzschild-Str. 2, D-85748 Garching bei M{\"u}nchen, Germany\\
$^4$Department of Physics, University of Athens, Panepistimiopolis, GR 15783 Zografos, Greece \\
$^5$Technische Universit{\"a}t M{\"u}nchen, James-Frank-Str. 1, D-85748  Garching bei M{\"u}nchen, Germany}

\begin{document}
\date{Received / Accepted}
\pagerange{\pageref{firstpage}--\pageref{lastpage}} \pubyear{2013}

\maketitle

\label{firstpage}

\begin{abstract}
The recent IceCube discovery of $0.1-1$ PeV neutrinos of astrophysical origin 
opens up a new era for high-energy astrophysics. Although there are 
various astrophysical candidate sources, a firm association of the detected neutrinos 
with one (or more) of them is still lacking. 
A recent analysis of plausible astrophysical counterparts within the error circles
of IceCube events showed that likely counterparts for nine of the IceCube neutrinos include mostly BL~Lacs,  among which Mrk~421.
Motivated by this result and  a previous independent analysis on the neutrino emission
from Mrk~421, we test the BL~Lac-neutrino connection in the context of
a specific theoretical model for BL~Lac emission.
We model the spectral energy 
distribution (SED) of the BL~Lacs selected as counterparts of the IceCube neutrinos using 
 a one-zone leptohadronic model and mostly nearly simultaneous data.
The neutrino flux for each BL~Lac is self-consistently calculated, 
using photon and proton distributions specifically derived for every individual source.
We find that the SEDs of the sample, although different in shape and flux,  are all well fitted by the model using reasonable parameter values.  Moreover, the model-predicted neutrino flux and energy for these sources are of the same order of magnitude as those
of the IceCube neutrinos. In two cases, namely Mrk~421 and H~1914-194,
we find a suggestively good agreement between the model prediction
and the detected neutrino flux. Our predictions for all the BL~Lacs of the sample
are in the range to 
be confirmed or disputed by IceCube in the next few years of data sampling.
\end{abstract} 
  

\section{Introduction}
% In 2012 the IceCube observatory reported
% the detection of two  neutrino events with energies of a few PeV,
% using  data obtained with the IC-79/IC-86 configuration, while
% a posterior analysis of the same dataset led to a sample
% of 28 events of likely astrophysical origin with energies 
% ranging from $30$~TeV to 2~PeV \citep{aartsen13}.
% An even more recent analysis of 3 years of combined data 
% not only raised the number from 28 to 37 events but also
% confirmed the hypothesis of an astrophysical
% origin at a 5.7$\sigma$ level \citep{aartsen14}.
 During the first three years of data sampling in full
configuration, the IceCube Neutrino Telescope based at the South Pole has 
registered a sample of very high-energy neutrinos (30~TeV -- 2~PeV) of astrophysical origin \citep{aartsen13, aartsen14}.
With a sample of 37 events the background only hypothesis with no astrophysical component has been rejected with 
a significance of more than 5~$\sigma$. A new era for high-energy astrophysics just started. 

Excluding, in this context, the possible
connection with dark matter \citep{cherryetal14, esmailietal14}, 
the IceCube discovery calls for astrophysical sites, where cosmic-ray acceleration at the ``knee'' energy scale and above
is efficiently at work, and the target, be it gas or  photons, is sufficiently dense for 
interactions with the cosmic rays.
% The importance of the {\sl cosmic} neutrino detection may be comparable
% to the discovery of high-energy $\gamma$-ray emission ($\gtrsim 100$~MeV) from active
% galactic nuclei (AGN), including both flat spectrum radio quasars (FSRQs) and BL~Lacs \citep{fichtel94}, which
% affected our understanding of the radiative processes
% at play.
% Similarly, the detection of cosmic neutrinos reveals
% the existence of astrophysical sites where cosmic ray acceleration
% is at work, and in this sense, is revolutionary.
A firm association between the detected neutrinos
and a specific class (or classes) of astrophysical sources
is still lacking, notwithstanding the numerous scenarios that have been
proposed so far. Models suggesting a pure Galactic origin of
the observed neutrino signal
can be strongly constrained by 
diffuse TeV-PeV $\gamma$-ray limits 
(see e.g.~\citealt{ahlersmurase13, joshi14} and references therein), given
that the production of very high energy (VHE) $\gamma$-rays is an inevitable
outcome of the hadronic interactions that lead to 
the neutrino production
in the first place.  
However, a Galactic component contributing to the IceCube neutrino flux cannot be excluded at the moment.
Proton-proton ($pp$) collisions in galaxy groups and/or star forming galaxies
have also been invoked to explain the diffuse neutrino flux (e.g. \citealt{loebwaxman06, muraseahlers13}).
Other extragalactic scenarios that include neutrino production through photohadronic ($\pg$) interactions
in active galactic nuclei (AGN) \citep{stecker91, mannheim95,
halzenzas97,rachen98, muecke99, atoyandermer01, muecke03} 
and gamma-ray bursts (GRBs) \citep{waxmanbahcall97, murase08, zhangkumar13, asanomeszaros14, reynoso14, petroetal14, winter14} have been 
extensively discussed in the literature. 

A commonly adopted approach for the calculation of the neutrino flux at Earth
(see e.g. \citealt{mannheimprotheroe01, baerwald13, kistler13, murase14})
is  the following: (i) a generic distribution of accelerated
protons in the source is assumed; (ii) a generic photon distribution that acts
as a target field for $\pg$ interactions is also assumed; (iii) the neutrino spectrum is then calculated; and
(iv) the neutrino flux is integrated over redshift assuming a luminosity function for the particular astrophysical source.
If many sources of the same type 
contribute to the observed flux, then the approach 
described above is the most relevant one.
A different approach is to 
focus on a few individual sources of the same class and derive 
the respective neutrino fluxes using proton and photon distributions that
are unique to each object \citep{muecke03, dpm14}. 
In this regard, information about the directions
of the neutrino events on the sky is important.
This approach may still lead to a study of the diffuse neutrino flux, 
from a certain class of sources, but with the caveat that it may 
not represent the total diffuse flux; the latter could
be arrived at only upon an estimate of the percentage of that class's contribution to the total.

% \maria{This approach may still be relevant to the diffuse neutrino
% flux from a certain class of sources.}
% % This may be of particular interest if the detected neutrinos originate
% % from  different classes of astrophysical sources, each of them having
% % its own characteristics. 
% \sd{In such a case, one can only generalize to a limited extent from 
% findings from members of a single class, and only upon an estimate 
% of the percentage of that class' contribution to the total flux.}

\cite{padovaniresconi14} -- henceforth PR14,  have recently searched
for plausible astrophysical counterparts 
within the error circles of IceCube neutrinos using high-energy $\gamma$-ray (GeV-TeV) catalogs.
Assuming that each neutrino event is associated with one astrophysical source and using
a model-independent method they derived the most probable counterparts for 9 out of the 18 neutrino events
of their sample. Interestingly, these include 8 BL~Lac objects, amongst which the nearest blazar, Mrk~421,
and two pulsar wind nebulae.  Although Flat Spectrum Radio Quasars (FSRQs) are believed to be more efficient PeV neutrino emitters
than BL~Lacs (e.g. \citealt{dermermurase14}; see, however, \cite{tavecchioetal14}), 
these do not appear in the list of most probable counterparts.
This is not necessarily related to the physics of neutrino emission from different
types of blazars but could be a consequence of 
the ``energetic'' test applied by PR14 (for more details, see \S 4 therein).
%We note also that \cite{sahu14} performed a similar analysis, yet restricted it to TeV emitting BL~Lacs.

Motivated by the aforementioned results, we investigate
the BL~Lac--PeV neutrino connection within a specific theoretical framework
for blazar emission where the $\gamma$-rays are of photohadronic origin. 
In contrast to the majority of works on neutrino emission
from blazars, we do not assume a generic SED but we fit instead the  multiwavelength (MW) photon spectra
   using the radiation of particles whose distributions have been accordingly modified
   by the cooling and escape processes. 
In our model the low-energy emission of the blazar SED 
is attributed to synchrotron radiation of
relativistic electrons, whereas the observed high-energy (GeV-TeV) emission is
the result of synchrotron radiation from pairs produced by charged pion
decays. Pions are the by-product of $\pg$ interactions
of co-accelerated protons with the internally produced synchrotron photons.
Our approach  allows us to identify the observed $\gamma$-ray emission as the emission from the $\pg$ component
and associate it with a very high-energy ($\sim2-20$~PeV) neutrino emission. Analytical expressions for 
the typical energies
of neutrinos and photons produced through photohadronic processes, the proton cooling rate due to 
photopion and photopair interactions, as well as the dependence of all aforementioned quantities on
observable quantities of blazar emission can be found in \cite{petromast14}.

First results of the model were  presented in \cite{dpm14} -- henceforth DPM14, where
the neutrino flux expected from the prototype blazar Mrk~421 in a low state was derived. 
%It is noteworthy that the prediction of the model about the muon neutrino flux 
The prediction of the model muon neutrino flux
was tantalizingly close to the IC-40 sensitivity limit 
calculated by \cite{tchernin2013} for Mrk~421 (see Fig.~1 in DPM14).
We extend our previous analysis to include those BL~Lacs 
identified as the most probable counterparts of the IceCube neutrinos by PR14 (see Table~4 therein).
Our aim is threefold: (i) to examine whether the working model (the so-called LH$\pi$ model, see DPM14) can fit 
the SED of the aforementioned blazars -- there was no guarantee before our attempts, since
we did not select {\sl a priori} sources with specific spectral characteristics; (ii) to calculate the neutrino flux 
relative to the $\gamma$-ray one, should acceptable fits be obtained, and (iii) 
to compare the model-derived neutrino fluxes with those calculated for each single neutrino event.
We show that Mrk~421 does not constitute a  unique case. Instead,  all 
BL~Lacs in our sample can be modelled by the leptohadronic model
and for most of them the model-predicted neutrino flux is close to the observed one, at least within the $3\sigma$ error bars.

This paper is structured as follows. We start with a short description of the theoretical 
model and the fitting method we used in Sect.~\ref{modelmethod}. In Sect.~\ref{case} we provide
general information about the observational datasets used for each BL~Lac. 
In Sect.~\ref{results} we present the combined MW photon and neutrino spectra
for each of the case studies, focusing on the ratio of the model-predicted neutrino luminosity to the
 $\gamma$-ray luminosity and its dependence on observable quantities. 
 We also comment on the model-derived energetics for each source. We discuss our results in 
 Sect.~\ref{discussion} and  conclude with a summary in Sect.~\ref{summary}.
Throughout this work we have adopted a cosmology 
with $\Omega_{\rm m}=0.3$, $\Omega_{\Lambda}=0.7$ and
$H_0=70$ km s$^{-1}$ Mpc$^{-1}$. For the attenuation of the VHE part of the
spectra  due to photon-photon pair production on the extragalactic background light (EBL) we used the
model of \cite{franceschini08}.

\section{Model and method}
\label{modelmethod}
\subsection{Model description}
\label{model}
The physical model we use has been described, in general terms, 
in \cite{DMPR2012} - henceforth DMPR12, and, in the context of 
modelling Mrk~421 in particular, in \cite{mastetal13} and DPM14. 
We consider a spherical blob of radius $\rb$, containing a 
tangled magnetic field of strength B and moving towards us 
with a Doppler factor $\delta$. 
Protons and electrons are accelerated by some mechanism whose 
details lie outside the immediate scope of this work, and 
are subsequently injected  isotropically in the volume of the blob with a constant rate.
Their interaction with the magnetic 
field and with secondary particles leads to the development 
of a 
%steady-state 
system where five stable particle populations 
are at work: protons, which lose energy by synchrotron radiation, 
Bethe-Heitler (pe) pair production, and photopion ($\pg$) interactions; 
electrons (including positrons), which lose energy by 
synchrotron radiation and inverse Compton scattering; 
photons, which gain and lose energy in a variety of ways; 
neutrons, which can escape almost unimpeded from the 
source region, with a small probability of photopion 
interactions; and neutrinos, which escape completely unimpeded.

The interplay of the processes governing the evolution 
of the energy distributions of those five populations is 
formulated with a set of time-dependent kinetic equations. 
Through them, energy is conserved in a self-consistent manner, 
since all the energy gained by a particle type has to come 
from an equal amount of energy lost by another particle type. 
To simultaneously solve the coupled kinetic equations for all 
particle types we use the time-dependent code described in DMPR12.
Photopion interactions are modelled using the results of the 
Monte Carlo event generator SOPHIA \citep{SOPHIA2000}, 
while Bethe--Heitler
pair production is similarly modelled with the Monte Carlo 
results of \cite{Protheroe1996} and \cite{mastetal05}. 
The only particles not modelled with kinetic equations are muons, 
pions and kaons. Their treatment is described in DPM14 and 
\cite{petroetal14} but is of negligible importance for the 
range of parameter values used in the present work.

The prime mover of all processes is the rate at which energy is injected 
in relativistic protons and electrons. This, in terms of 
compactnesses,  is expressed as:
\eqb
\ell_{\rm i, inj}={{L'_{\rm i, inj} \sth}\over{4\pi R m_{\rm i} c^3}},
\label{lpinj}
\eqe
where i denotes protons or electrons (i=e,p),  
$\sigma_{\rm T}$ is the Thomson cross section, and $L'_{\rm i, inj}$ denotes the injection luminosity as measured in 
the rest frame of the emitting region.
In principle, the radius $R$ and Doppler factor $\delta$ can be
constrained using variability arguments. However, we 
prefer to treat them as free parameters, since 
information for high-energy variability is not available for
all sources in the sample.
The escape times for protons and electrons are assumed to be 
the same ($\tpesc=\teesc$), and complete the set of free parameters 
dependent on the geometry and kinematics of the blob. 
What remains to be defined are the shapes of the proton and primary 
electron injection functions, which are described as power-laws with index
$s_{\rm i}$ extending from $\gamma_{\rm i, \min}$ to $\gamma_{\rm i, \max}$.
In some cases, modelling of the SED requires the primary electron distribution to be
a broken power-law, with the break energy $\gamma_{\rm e, br}$ and the 
 power-law index above the break $s_{\rm e, h}$ being two additional free parameters. 

 As long as the escaping protons and neutrons are energetic enough, they
are susceptible to photopion 
interactions with ambient photons in Galactic and intergalactic 
space, such as the cosmic microwave  and infrared backgrounds,  producing 
additional high-energy neutrinos \citep{stecker73}. 
Neutrons also rapidly decay into protons 
\citep{sikoraetal87, kirkmast89, giovanonikazanas90, atoyandermer01}, 
leading also to neutrino production.
% However, as we have shown in DPM14, in the case of Mrk~421 at least 
% they have a negligible impact on the total neutrino luminosity. 
In this work we focus on the neutrino emission 
from the objects themselves, and will not consider any such 
additional contributions from escaping high-energy nucleons.

The reader should note that we placed our emphasis on fitting the MW 
data with a variant of the leptohadronic model where the $\gamma$-ray emission is 
attributed to synchrotron radiation of the secondaries arising from charged pion decay.
Clearly, this is not the only way of fitting the SED, 
as it is possible to fit the $\gamma$-rays with proton synchrotron radiation (e.g. \citealt{aharonian00}). In that case,
however, the model produces a very low neutrino flux at $\sim$EeV energies (see LHs model in DPM14), which makes it 
irrelevant to the  IceCube TeV-PeV detections. 
Moreover, scenarios where the high-energy blazar emission is the result
of $\gamma$-rays produced through $pp$ collisions within the blazar jet, constitute another alternative\footnote{
To our knowledge, direct application of such models to MW observations of blazars is still lacking in the literature.}. Such models
would require, however, uncomfortably high densities of non-relativistic (cold) protons to be present 
in the blazar jet (see e.g.~\citealt{atoyandermer04} and references therein). 
Last but not least, there is always the possibility of assuming a 
purely leptonic model \citep{boettcherreimer13} which, however, does not produce any neutrinos. 
% In summary, 
% our aim was twofold: to examine whether the LH$\pi$ model can fit 
% the SED of the aforementioned blazars - of which there was no guarantee before our attempts, 
% and to calculate the neutrino flux relative to the $\gamma$-ray one, should acceptable fits be obtained.
\subsection{Fitting method}
\begin{figure}
 \centering
\includegraphics[width=0.46\textwidth]{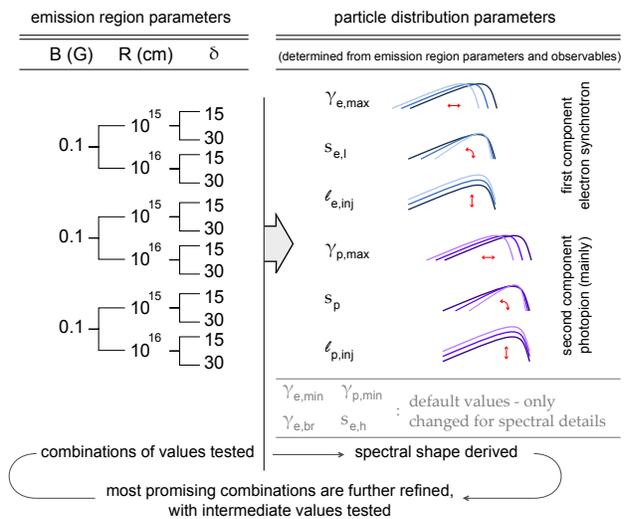}
 \caption{Flow diagram illustrating the steps of the fitting method.}
 \label{fitting}
\end{figure}
The multi-parameter nature of the problem (11-13 free parameters; see Sect.~\ref{results}) and
the long simulation time, which is required for every parameter set (of the order of $\sim0.5$ d),
prevents us from a blind search of the whole available parameter space. 
Moreover, the DMPR12 code in its current form has not been developed for parallel computing.
Yet, the application of the model to actual observations is still viable (e.g. \citealt{mastetal13, petro14})
thanks to the insight we gain from an analytical treatment of the problem.
Thus, instead of automatically scanning the parameter space,
we start the fitting procedure with a set of values, which are physically motivated (see below).
Then, we perform a series of numerical simulations
with parameter values lying close to the initial parameter set,
until a reasonably good fit to the SED is obtained. The term ``reasonably good'' fit
refers to a model SED that passes through most observational points. This is not necessarily the best fit
in the strict sense, i.e. the one with the lowest $\chi^2$ value. However, 
the main results of our study, such as the $\gamma$-ray/neutrino connection, 
are not affected by the exact value of the $\chi^2$ statistics. 

The steps of the fitting method
are illustrated in Fig.~\ref{fitting}.
The parameters of the problem can be divided into two groups; those related to the properties of the
emission region, i.e. $B$, $\rb$ and $\delta$ and those
related to the primary particle distributions (see two columns in Fig.~\ref{fitting}). 
As benchmark values for the magnetic field we chose $B=0.1, 1, 10$~G. These cover the range 
of low-to-moderate magnetic fields. Higher values, e.g. $B\gtrsim 20$~G, were not included to the benchmark set, since
they tend to favour proton synchrotron radiation against photohadronic processes (for proton synchrotron models, see
\citealt{aharonian00, muecke03, boettcherreimer13, dpm14, cerrutietal14}).
For each value of $B$, we then adopted  $\rb =10^{15}$~cm and $10^{16}$~cm and the 
set of benchmark values is completed with $\delta=15$ and $30$. 
We did not search in the first place for fits having 
even higher values of $\delta$ and $\rb$, since they would decrease 
the optical depth ($f_{\pg}$) for photopion interactions and for neutrino production.
If the low-energy hump of the SED has peak spectral index $\beta$, peak frequency $\nu_{\rm s}$
and total luminosity $L_{\rm syn}$ the optical depth for neutrino production is given by (e.g. PM15)
\eqb
f_{\pg} (\gamma_{\rm p}) 
\simeq 4.4 \times 10^{-3} \frac{L_{\rm syn, 45} \lambda(\beta, \nu_{\rm s}) }{R_{15} \delta_1^3 \nu_{\rm s, 16} (1+z)}
\left(\frac{2 \gamma_{\rm p}}{\gamma_{\rm p}^{(\pg)}}\right)^{\beta}, \gamma_{\rm p} > \gamma_{\rm p}^{(\pg)}\frac{\nu_{\rm s}}{2 \nu_{\min}},
\label{eq1}
\eqe
where we introduced the notation $q_{x}=q/10^x$ in cgs units,
$\lambda$ is a function of $\beta$ and $\nu_{\rm s}$ (see equations~(23) and (27) in PM15), $z$ is the redshift of the source, 
$\nu_{\min}$ is the low cutoff frequency of the low-energy spectrum (for a continuous power-law), $\gamma_{\rm p}$
is the proton Lorentz factor and $\gamma_{\rm p}^{(\pg)}$ is the minimum Lorentz factor that satisfies the threshold
condition for $\pg$ interactions with the peak energy photons. This is given by (for details, see PM15)
\eqb
\gamma_{\rm p}^{(\pg)} \simeq 3.5 \times 10^7 \delta_1 \nu_{\rm s, 16}^{-1} (1+z)^{-1}.
\label{eq2}
\eqe
Protons with the aforementioned Lorentz factor lead to the production of neutrinos with energy, as measured in the observer's frame, given by
\eqb
\epsilon_{\nu} \simeq 17.5 \ {\rm PeV} \delta_1^2(1+z)^{-2} \nu_{\rm s, 16}^{-1}.
\label{Ev}
\eqe
In most cases, the above expressions is a safe estimate of the peak energy of the neutrino spectrum.

We use equation~(\ref{eq2}) as the minimum requirement for the maximum proton energy of 
the proton distribution, i.e. $\gamma_{\rm p, \max} = \kappa \gamma_{\rm p}^{(\pg)}$ with $\kappa$ a numerical factor of $\sim 2-5$. 
On the other hand, the maximum (or break) Lorentz factor of the primary electron distribution is related
to the peak frequency of the low-energy hump as
\eqb
\gamma_{\rm e, \max/\rm br} \simeq 6 \times 10^4 \left(\frac{\nu_{\rm s, 16} (1+z)}{\delta_1 B_{-1}}\right)^{1/2}.
\label{eq3}
\eqe
For the lower cutoffs of the proton and primary electron injection functions we use as default values
$\gamma_{\rm e,p \min}=1$ (grey colored parameters in Fig.~\ref{fitting}).
If we find large deviations of the model SED from the observations 
at $\nu < \nu_{\rm s}$, we search for $\gamma_{\rm e, \min} > 1$. 
We search for different $\gamma_{\rm p,\min}$ than the default one 
only if the proton distribution has $s_{\rm p} \ge 2$ and extension to $\gamma_{\rm p, \min}=1$
leads to extremely high injection luminosities. We note, however, that this is not a typical case.

The power-law index of the primary electron distribution ($s_{\rm e}$ or $s_{\rm e, h}$) is directly
related to the observed spectral index of the low-energy component in the SED. 
However, it is not straightforward to relate $s_{\rm p}$ with the observed photon
index, e.g. in the Fermi/LAT regime, where the emission from secondaries produced in photohadronic interactions dominates.
Thus, as a first estimate we use $s_{\rm p} \simeq s_{\rm e}$.

The injection luminosity of primary electrons is also directly related to the observed luminosity of the low-energy hump
as  (e.g. \citealt{mastkirk97})
\eqb
\leinj = 2\times10^{-4}\frac{L_{\rm syn, 45}}{R_{15} \delta_1^4}.
\label{eq4}
\eqe
To derive a first estimate of the proton injection luminosity we assume that the $\gamma$-ray ($> 10$~GeV)
luminosity is totally explained by the synchrotron emission of $\pg$ pairs:
%Using the relations
\eqb
\label{eq00}
L_{\gamma} & \approx & \delta^4 \epsilon'_{\gamma} L'_{\pg}(\epsilon'_{\gamma})\big|_{{\rm peak}} \\
\epsilon'_{\gamma} L'_{\pg}(\epsilon'_{\gamma}) &  \approx & \frac{1}{8} 
\label{eq01}
f_{\pg}\left(\epsilon'_{\rm p}\right)\epsilon'_{\rm p} L'_{\rm p, inj}\left(\epsilon'_{\rm p} \right) \approx \frac{1}{8} 
f_{\pg}\left(\epsilon'_{\rm p}\right) L'_{\rm p, inj},
\label{eq02}
\eqe
where $f_{\pg}$ is given by equation~(\ref{eq1}). The factor $1/8$ in equation~(\ref{eq01}) comes
 from the assumption that half of the produced pions produce electrons/positron pairs, and each of them gains $\sim 1/4$ of the parent
 proton energy:
\eqb
\label{eq03}
\epsilon'_{\rm p} &  = & \gamma_{\rm p} \mpr c^2.
\eqe
The characteristic photon energy is given by
\eqb
\label{eq04}
\epsilon'_{\gamma} & \simeq & \mel c^2 \frac{B}{B_{\rm cr}} \gamma_{\rm e, \pg}^2,
\eqe
where 
\eqb
\gamma_{\rm e, \pg} & \simeq & \kappa_{\pg} \frac{\mpr}{4 \mel} \gamma_{\rm p}, \ \kappa_{\pg}=0.2 
\label{eq05}
\eqe
Using equations~(\ref{eq00})-(\ref{eq05}), (\ref{lpinj}), and (\ref{eq1}) we derive a rough estimate of the proton injection compactness:
\eqb
\lpinj \simeq 2.3 \times 10^{-3}
\frac{L_{\gamma, 46}}{L_{\rm syn, 45}}\frac{\nu_{\rm s,16}(1+z)}{\delta_1 2^{\beta} \lambda(\beta, \nu_{\rm s})}.
\label{eq5}
\eqe
Summarizing, for each set of benchmark values for $B, \rb$ and $\delta$ and equipped with the analytical estimates of
equations~(\ref{eq1})-(\ref{eq5}) 
and observables, we may significantly reduce the parameter space we need to scan (Fig.~\ref{fitting}).
\section{Case studies}
\label{case}
Our modelling targets are BL Lacs that are   spatially and energetically correlated 
with neutrino detections. BL~Lacs
are known for their variable emission 
at almost all energies. In high-states they usually exhibit major
flares, especially in X-rays and $\gamma$-rays. 
Modelling of their broadband emission in low and high states
requires a different parameterization. For example,
a high-flux state might  need a
%higher injection luminosity of
%radiating particles or/and a 
stronger magnetic field (for modelling of blazar flares, see \citealt{mastkirk97, coppiaharonian99, sikoraetal01, krawczynski02}),
which in turn would affect the resulting $\gamma$-ray and neutrino
fluxes.
Thus, it is important to build an SED using simultaneous data or,  if this is not possible, using
MW observations that have been obtained at least within the same year.

\begin{figure}
\centering
\includegraphics[width=0.45\textwidth]{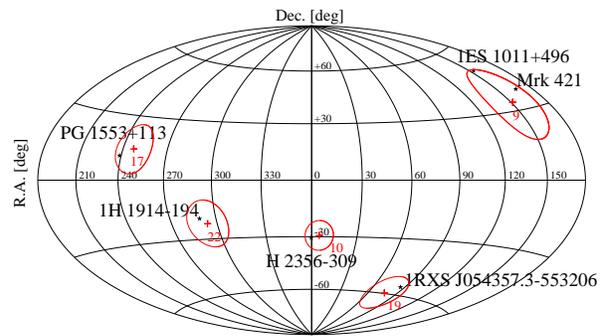}
\caption{Sky map in equatorial coordinates of the five IceCube neutrino events (crosses) that have as probable astrophysical counterparts
the six  BL~Lac sources (stars) mentioned in text. The red circles correspond to the median angular
error (in degrees) for each neutrino event (see Table~\ref{tab-00}).}
\label{map}
\end{figure}

Out of the 8 BL Lacs reported by 
PR14, two have unknown redshifts: SUMSS J014347-584550
and PMN J0816-1311, which are the probable counterparts of neutrinos 20 and 27, respectively. Since redshift is a crucial 
parameter in determining the luminosity of the emitting regions 
and, thus, the interplay of processes at work there, we have 
chosen to exclude those two BL Lacs from this study.  
Therefore, we are left with six targets, which are:
\begin{itemize}
 \item Mrk~421 at $z=0.031$ \citep{Sbarufatti2005_Mrk421z}. 
This is the closest high-frequency peaked blazar (HBL) and the target of many MW campaigns. 
  Mrk~421 was also the first source which our model (see Sect.~\ref{model}) was applied to and the
main focus of an earlier work (DPM14). 
We first quote the results of DPM14, that
used the simultaneous X-ray (RXTE) and TeV $\gamma$-ray  (Whipple and HEGRA)
 data of the 2001 MW campaign  that lasted 6 days \citep{Fossati2008}. 
 In particular,  DPM14 fitted the SED of March 22nd/23rd 2001 that corresponds to a pre-flare state.
 During this period no GeV observations were available. Thus,
 the Fermi/LAT data  during the period of the IC-40 configuration \citep{Abdo2011} were included for comparison reasons only.
As a second step, we use the MW observations of \cite{Abdo2011} together with the 
neutrino event ID 9 in order to build a {\sl hybrid} SED. 
To avoid repetition and to emphasize the prediction of DPM14 regarding the muon neutrino flux,
we do not attempt to 
fit the new dataset. Instead, we use the same parameters as in DPM14, apart from the Doppler factor, and
compare the model results against the \cite{Abdo2011} observations and the neutrino flux for the event ID 9.
  \item PG~1553+113 at $z=0.4$ \citep{aleksic12}.  This is not only the most distant source of this sample, but
 also belongs to the class of most distant TeV blazars in general. Its redshift has not been firmly measured, 
 yet the value adopted here can be considered
 as a safe estimate (for  different techniques and results, see \citealt{aleksic12}). 
 Similarly to 1ES~1011+496 below, it has been only recently detected in VHE $\gamma$-rays by
 both H.E.S.S. \citep{aharonian06} and MAGIC \citep{albert07b}. 
The data from the second year Fermi-LAT 
 Sources Catalog at $\sim 1$ GeV (2FGL\_lc) \footnote{http://www.asdc.asi.it/fermi2fgl/} indicate a variable high-energy flux. 
 The 2005-2009 MAGIC observations also suggest variability at the VHE part of the spectrum (see Fig.~1 in \citealt{aleksic12}).
 In 2005 the source exhibited large flux variations in X-rays, yet there were no simultaneous data, neither at  GeV nor at TeV energies.
We adopted the data shown in Fig.~6 of \cite{aleksic12}.
For the SED fitting we used, in particular, 
the X-ray data that correspond to the intermediate flux
level of 2005 along with the time-averaged 2008-2009 Fermi/LAT data and the MAGIC observations
for the periods
2005-2006 and 2007-2009. We 
include in our figures the high state X-ray data of 2005 only for illustrative reasons. 
In principle, one could try to fit the time-averaged $\gamma$-ray data
together with the 2005 high state in X-rays. 
However, we find  less  plausible a scenario where the 2005 X-ray high-state is not accompanied by an increased
$\gamma$-ray flux but, instead, is related to the average $\gamma$-ray flux level. There are two reasons for this:
the $\gamma$-ray and X-ray emission in blazars are usually correlated  (e.g. for Mrk~421, 
see \citealt{Fossati2008}) and the $\gamma$-ray emission of PG~1553+113 is itself
variable (see e.g. \citealt{aharonian06}).
\item 1ES 1011+496 at $z=0.212$ \citep{albert07}. This is an HBL source at indermediate
 redshift, that has been detected for the first time in TeV $\gamma$-rays in 2007 with MAGIC \citep{albert07}. The 
 discovery was the result of a follow up observation of the optical high state reported by the Tuorla blazar monitoring program\footnote{http://users.utu.fi/kani/1m}.
 The previous non-detections in TeV $\gamma$-rays suggest that 1ES~1011+496 is variable
 both in optical and VHE $\gamma$-rays. Its X-ray emission is described by a
 steep spectrum in both low and high states (see \citealt{reinthal12} and references therein).
 Because of the variability observed across the MW spectrum, we chose to apply our model to simultaneous data, if possible.
 We therefore used data from the first MW campaign in 2008 as presented in  \cite{reinthal12}, despite the fact
 that they are characterized as preliminary.
 We note that the MAGIC data presented in \cite{reinthal12} have been de-absorbed using
 the EBL model of \cite{kneiske10}. Using the same model we retrieved the observed MAGIC spectrum, which
 we show in the respective figure.
 We also took into account the Fermi/LAT data from the first (1FGL)  \citep{abdo10} 
 and second (2FGL)  \citep{nolan12} catalogs. In the respective figures we include 2FGL\_lc data, in order to show the variability in the GeV regime.
  \item H~2356-309 at $z=0.165$ \citep{falomo91}. This is one of the brightest HBLs in X-rays, which led
 to its early X-ray detection with UHURU \citep{forman78}. 
 Its X-ray emission is variable both in flux and in spectral shape (for more information, see \citealt{abramowski10}).
  TeV $\gamma$-ray emission from H 2356-309 was observed for the first time in 2004 by H.E.S.S. \citep{aharonian06b}.
 Using the SED builder tool\footnote{http://www.asdc.asi.it/SED} of the ASI Science Data Centre (ASDC)
 we chose those X-ray observations that fall within the period
 2004-2007 of H.E.S.S. observations. We also included data from the 1FGL and 2FGL catalogs, although
 they refer to the period 2008-2010, as well as hard X-ray data obtained by Swift/BAT between 2004 and 2010.
 Finally, we over-plotted the historical X-ray high state observed with BeppoSaX, for comparison
 reasons.
  \item 1H~1914-194 at $z=0.137$ \citep{Carangelo2003_1H1914z}. This is the first BL~Lac of the sample that has not yet
  been detected in TeV $\gamma$-rays, and the least well covered in different energy bands among the sources of this sample.
   In order to apply our model we constructed a SED using available data from the ASDC, which
   is comprised mainly of NED archival data. Given the quality of the available dataset, our main focus was 
  to describe as  well as possible the Fermi/LAT data, while roughly describing the observations in other energy bands.
 \item \rxs \ at $z=0.374$ \citep{Stadnik2014_1RXSz}. This is the second most distant object of the sample, after blazar PG~1553+113.
It was first detected and classified as a BL~Lac by ROSAT \citep{Fischer1998AN}. 
Following radio observations by \cite{Anderson2009} it was classified as an HBL, 
and it was soon thereafter detected in $\gamma$-rays by Fermi/LAT \citep{abdo10}. 
It remains still undetected in TeV energies. Its redshift was 
estimated by \cite{Stadnik2014_1RXSz}, based on the assumption that the flux of the host elliptical galaxies of 
BL Lacs can be used as a standard candle, and also by \cite{Pita2014}, based on spectroscopy on a weak 
Ca II H/K doublet and a Na~I absorption line. While the \cite{Pita2014} estimate 
($z=0.237$) is somewhat lower than the  \cite{Stadnik2014_1RXSz} one, adopting it would 
not significantly alter out fit.
Using the SED builder tool of ASDC we combined infrared (WISE), X-ray (Swift/XRT and XMM-Newton) and $\gamma$-ray 
(Fermi/LAT) observations that were obtained within the same year (2010).
We also included older XMM-Newton observations performed in 2005 and 2007, for comparison reasons. Finally, the 
2FGL\_lc data illustrate the variability in the GeV energy band.
\end{itemize}
Before closing this section and for completeness reasons, we 
summarize in Table~\ref{tab-00}, 
the coordinates of the five IceCube neutrinos and the six most probable blazar counterparts, as well as
their angular offset. We also included the median angular error for each neutrino detection and the
respective detection time in Modified Julian Days (MJD)\footnote{We note that all
data listed in Table~\ref{tab-00} are adopted from Tables 1 and 2 in PR14.}. A sky map (in equatorial coordinates) 
of the five IceCube neutrinos and their respective counterparts is shown in Figure~\ref{map}.

\begin{table*}
\caption{List of the most probable counterparts of selected high-energy IceCube neutrinos used in our modelling.}
\begin{tabular}{ccc ccl ccc}
 \hline
IceCube ID & Ra(deg) & Dec(deg) & Median angular error (deg) & Time (MJD) &  Counterpart &  Ra(deg) & Dec(deg) & Angular offset (deg) \\
 \hline
 9 & 151.25 &   33.6 & 16.5 &  55685.6629638 & Mrk~421 &  166.08 & 38.2 & 12.8 \\
 \multicolumn{5}{c}{\phantom{a}} & 1ES~1011+496 & 155.77 & 49.4 & 15.9 \\
 10 & 5.00 & -29.4 & 8.1 &  55695.2730442 & H~2356-309 & 359.78 &  -30.6 & 4.7 \\
 17 & 247.40 &  14.5 &  11.6 &  55800.3755444 & PG~1553+113 &  238.93& 11.2  & 8.9 \\
 19 & 76.90  & -59.7 &  9.7 &  55925.7958570 & 1RXS~J054357.3-553206 &  85.98 & -55.5& 6.4\\
 22 & 293.70 & -22.1 &	12.1 & 55941.9757760 & 1H~1914-194&  289.44 & -19.4 &   4.8\\
 \hline
\end{tabular}
 \label{tab-00}
\end{table*}

\begin{figure*}
 \centering
\includegraphics[width=0.45\textwidth]{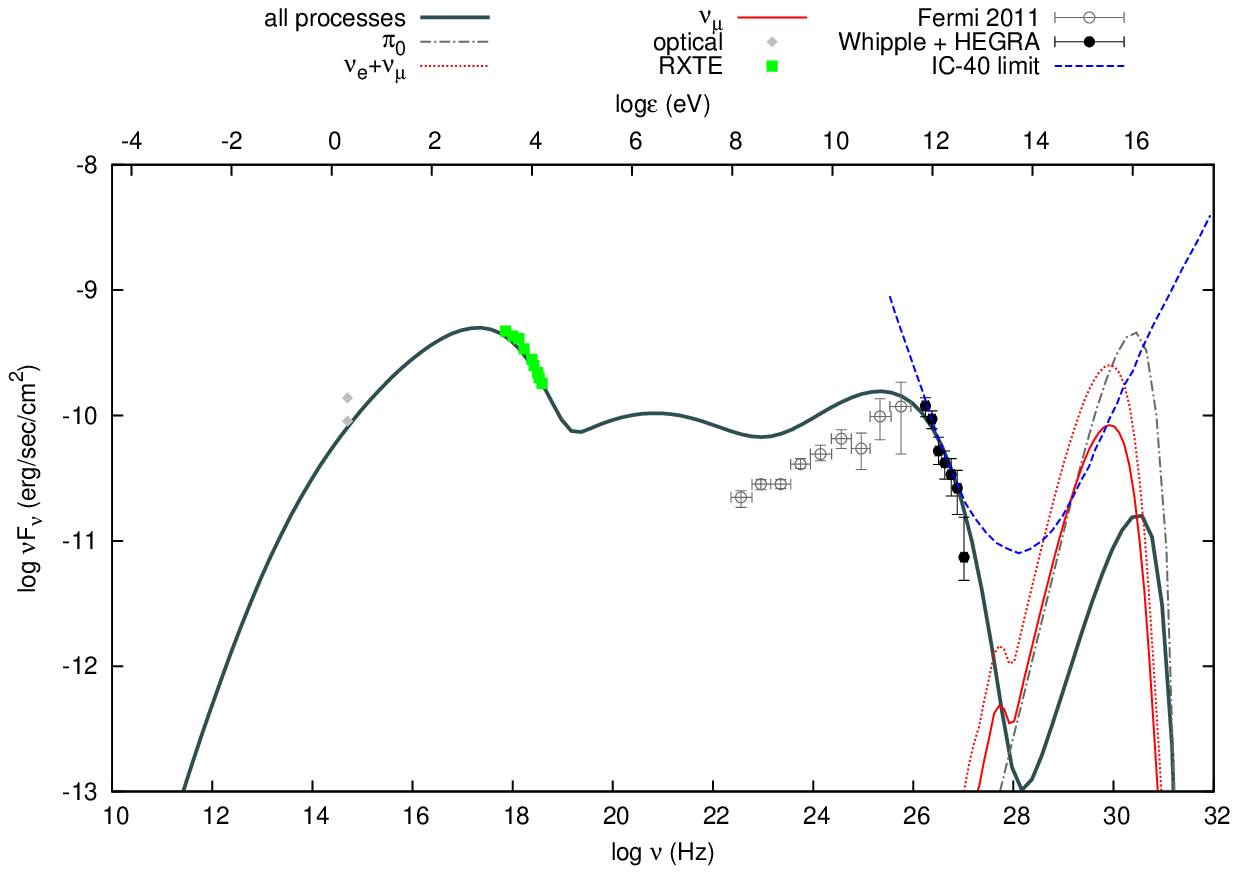}
\includegraphics[width=0.46\textwidth]{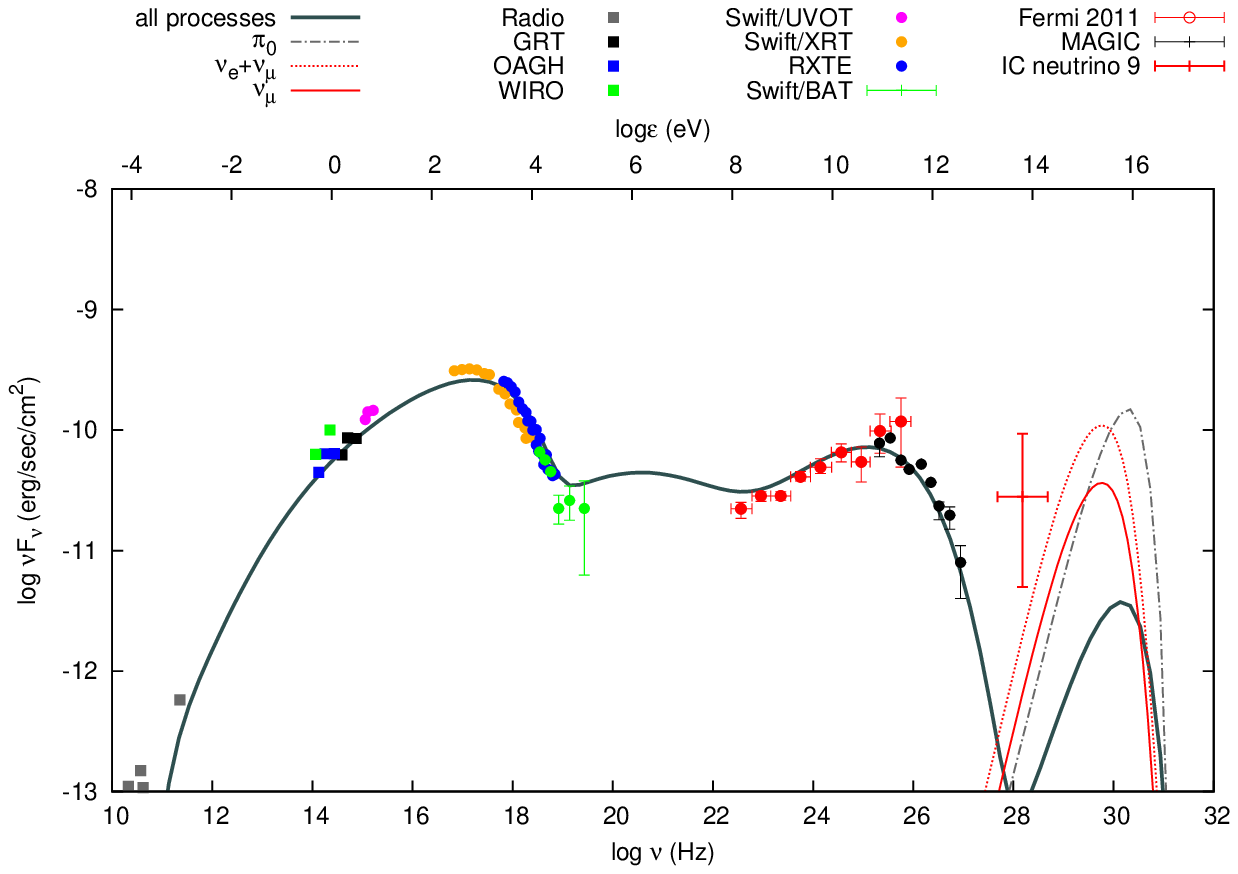}
 \caption{Left panel: SED of blazar Mrk~421 as modelled in DPM14. The Fermi/LAT points are shown only for illustrative reasons, as they
 were not included in the fit of the original paper DPM14. Right panel: SED of Mrk~421 averaged over the period
 19 January 2009 - 1 June 2009 (Abdo et al. 2011); all data points are from Fig.~4 in Abdo et al. (2011).
 The model SED (black line) and neutrino spectra (red lines) 
 are obtained for the same parameters as in DPM14, except for the Doppler factor, i.e. $\delta=20$.
 The neutrino event ID 9 is also shown. In both panels, the model spectra are not corrected for absorption on the EBL. 
 For comparison reasons, the unattenuated $\gamma$-ray emission from the $\pi^0$ decay is over-plotted with dash-dotted grey lines.}
 \label{mrk421}
\end{figure*}
\section{Results}
\label{results}
We first present the results of our SED modelling along with the predicted neutrino flux
for each source, i.e. we build model {\sl hybrid} SEDs. The parameter values
derived for each source are summarized in Table~\ref{tab-0}. Various aspects of the 
adopted theoretical framework were investigated in detail in a 
recent work (\citealt{petromast14} -- hereafter PM15). For details and analytical expressions
that give insight to the results that follow, we refer the reader to PM15. 
In the last two subsections we discuss in more detail the neutrino emission and the energetics 
for each source.
%we have investigated
% in detail various aspects of the theoretical model we adopt for modelling the SEDs,
% and presented analytical expressions that  are useful, while commenting
% on the fitting results.
\begin{table*}
 \caption{Parameter values for the SED modelling of blazars shown in Table~4 in PR14. The ordering of presentation is the same as
 in Sect.~\ref{results}.
 The parenthesis below the name of each source encloses the ID of the associated neutrino event. The double horizontal line
 separates parameters used as an input to the numerical code (upper table) from
 those that are derived from it (lower table).}
 \begin{threeparttable}
 \begin{tabular}{cccc ccc}
  \hline\hline
  Parameter & Mrk~421  &  PG~1553+113 &1ES~1011+496  & H~2356-309 & 1H~1914-194 & 1RXS~J054357.3-553206 \\
            & (ID 9)   & (ID 17)      &   (ID 9)    &    (ID 10)    &  (ID 22)    & (ID 19)\\
  \hline\hline
  z & 0.031 & 0.4 & 0.212&  0.165 & 0.137 & 0.374 \\
  B (G) & 5 &  0.05 & 0.1 & 5 & 5 & 0.1\\
  $\rb$ (cm) & $3\times 10^{15}$&  $2\times 10^{17}$ & $3\times 10^{16}$ & $3\times 10^{15}$ & $3 \times 10^{15}$ & $3\times 10^{16}$\\
  $\delta$ & 26.5 &  30 & 33 & 30 & 18 & 31\\
  \hline
  $\gamma_{\rm e, \min}$ & 1 & 1 & 1 & $1.2 \times 10^3$ & $1.2\times10$ & 1 \\
  $\gamma_{\rm e, \max}$ & $8\times 10^4$& $2\times 10^6$ & $10^5$\tnote{a} &  $10^5$  & $10^5$ & $1.2 \times 10^5$\tnote{a} \\
  $\gamma_{\rm e, br}$ &  -- &  $2.5 \times 10^5$ & -- & $2\times 10^{4}$ &  $10^3$ & -- \\
  $s_{\rm e, l}$ & 1.2 & 1.7 & 2.0 &  1.7 &  2.0 & 1.7\\
  $s_{\rm e, h} $ & -- & 3.7 &  -- & 2.0 & 3.0  &-- \\
  $\ell_{\rm e, inj}$ & $3.2 \times 10^{-5}$ &  $2.5\times 10^{-5}$ &  $4\times 10^{-5}$ &$8\times 10^{-6}$ & $6\times 10^{-5}$ & $2.5 \times 10^{-5}$\\ 
  \hline
  $\gamma_{\rm p, \min}$ & 1 &  $10^3$ & 1 & 1 & 1 & 1 \\
  $\gamma_{\rm p, \max}$ & $3.2\times 10^6$ & $6\times 10^6$ & $1.2\times 10^7$& $10^7$ &  $2\times 10^6$ & $1.2 \times 10^7$\\
  $s_{\rm p}$ & 1.3 & 2.0 & 2.0 & 2.0 & $1.7$ & 2.0\\
  $\ell_{\rm p, inj}$& $2\times 10^{-3}$ & $3\times 10^{-4}$ & $4\times 10^{-3}$ &  $2.5\times 10^{-3}$ & $10^{-2}$ & $8 \times 10^{-4}$ \\
   \hline 
  \hline
   $L'_{\rm e, inj}$~(erg/s)\tnote{b} & $4.4\times 10^{40}$  &  $2.3\times 10^{42}$ & $5.6 \times 10^{41}$ & $1.1\times 10^{40}$ & $8.3\times 10^{40}$& $3.5 \times 10^{41}$  \\
    $L'_{\rm p, inj}$~(erg/s)\tnote{b} & $5.1 \times 10^{45}$& $5.1 \times 10^{46}$ &  $10^{47}$ & $6.4 \times 10^{45}$ &
    $2.5\times 10^{46}$& $2 \times 10^{46}$\\
  $L_{\gamma, \rm TeV}$~(erg/s)\tnote{c}& $1.7\times10^{45}$ & $7.9 \times 10^{46}$ & $5\times 10^{45}$ &  $5\times10^{44}$ & $ 10^{45}$ & $6.3\times 10^{45}$ \\  
  $L_{\nu}$~(erg/s)\tnote{d} & $2.5\times 10^{45}$ & $8.1 \times 10^{45}$ & $2.5\times 10^{45}$ &  $4\times 10^{44}$ & $2 \times 10^{45}$ & $6.3\times 10^{44}$ \\ 
  $\ygn$\tnote{e} &1.5 & 0.1 &  0.5\tnote{f}& 0.8& $2.0$ & 0.1\\
  $f_{\pg}$\tnote{g} & $2\times 10^{-5}$ & $6\times 10^{-6}$ & $1.3\times 10^{-6}$ &  $3.1\times 10^{-6}$ & $1.2\times10^{-5}$ & $1.1\times 10^{-5}$\\
  \hline
 \end{tabular}
  \begin{tablenotes}
 \item[a] An exponential cutoff of the form $e^{-\gamma/\gamma_{\rm e, \max}}$ was used.
 \item[b] Proton and electron injection luminosities are given in the comoving frame.
 \item[c] Integrated $0.01-1$~TeV $\gamma$-ray luminosity of the model SED. Basically, this coincides with the observed value.
 \item[d] Observed total neutrino luminosity, i.e. $(\nu_{\mu} + \overline{\nu}_{\mu}) + (\nu_{\rm e} + \overline{\nu}_{\rm e})$.
 \item[e] $\ygn$ is defined in equation~(\ref{Ygn}).
 \item[f] This is the value derived without including in our modelling
 the upper limit in hard X-rays (see Sect.~\ref{1es}). For this reason, it should be considered only
 as an upper limit.
 \item[g] Estimate of the $\pg$ optical depth (or efficiency). We define it as $f_{\pg} \simeq 8  L'_{\gamma, \rm TeV}/L'_{\rm p, inj}$,
 where $L'_{\gamma, \rm TeV}$ is the integrated $0.01-1$~TeV $\gamma$-ray luminosity as measured in the comoving frame.  The values are only an upper limit because of the assumption that the observed $\gamma$-ray luminosity is totally explained
 by  $\pg$ interactions.
 \end{tablenotes}
 \end{threeparttable}
\label{tab-0}
 \end{table*}
\subsection{SED modelling}
\subsubsection{Mrk~421}
As already pointed out in the introduction, the most intriguing result of the
leptohadronic modelling of Mrk~421 was the predicted muon neutrino flux, which
was close to the IC-40 sensitivity limit calculated for the particular source by \cite{tchernin2013}.
This is shown in Fig.~\ref{mrk421} (left panel). 
We note that in the original paper DPM14, whose results we quote here, the Fermi/LAT data \citep{Abdo2011}
shown as grey open symbols in Fig.~\ref{mrk421} were not included in the fit, since they were not simultaneous with the rest of the observations.
The total and muon neutrino spectra are plotted with dotted and solid red lines, respectively.
We use the same convention in all figures that follow, unless stated otherwise. 
The bump that appears two orders of magnitude below
the peak of the neutrino spectrum in energy originates from neutron decay; we refer the reader to 
DPM14 for more details. 

According to PR14, Mrk~421 is the most probable counterpart of the corresponding
IceCube event (ID 9). This is now included in the right panel of Fig.~\ref{mrk421}, where
the time-averaged SED during  the
MW campaign of January 2009 - June 2009 \citep{Abdo2011} is also shown\footnote{All data points are taken from Fig.~4 in \cite{Abdo2011}.}.
As we have already pointed out in Sect.~\ref{case}, in order to avoid repetition, 
we did not attempt to fit the new data set of \cite{Abdo2011}. 
Thus, the model photon (black line) and neutrino (red line) spectra 
depicted in the right panel of Fig.~\ref{mrk421}, 
were obtained for the same parameters used in DPM14 (see also Table~\ref{tab-0}), except for the Doppler factor, which was 
here chosen to be $\delta=20$. The fit to the 2009 SED is surprisinly good, if we
consider the fact that we used the same parameter set as in DPM14 with only a small modification in the value of the Doppler factor.

We note that the MW photon spectra shown in both panels have not been corrected for EBL absorption in order to demonstrate 
the relative peak energy and flux of the neutrino and $\pi^0$ components. The $\gamma$-ray emission from the $\pi^0$
decay is shown in both panels as a bump in the photon spectrum, peaking at $\sim 10^{30}$~Hz. Although
the model prediction for the ratio of the $\pi^0$ $\gamma$-ray luminosity to the total neutrino luminosity is roughly two (see e.g. DMPR12), 
the $\pi^0$ component shown in Fig.~\ref{mrk421} is suppressed because of internal photon-photon absorption.
To  demonstrate better the effect of the latter, we over-plotted the $\pi^0$ component that is
obtained when photon-photon absorption  is omitted (dash-dotted grey lines). In the figures that follow,
we will display, for clarity reasons, the $\pi^0$ component only before its attenuation by the internal synchrotron and EBL photons.
%This also explains the absence of the $\pi^0$
%component in the figures that follow.}

\cite{Abdo2011}, whose data we use here,  have also presented a hadronic fit to the SED. Thus,
a qualitative comparison between the two models is worthwhile. Both models are similar in that they require a compact emitting region and 
hard injection energy spectra of protons and primary electrons.  Morevoer, 
the $\gamma$-ray emission from MeV to TeV energies, in both models, has a significant
contribution from the hadronic component. However, the models differ in several aspects because of: (i) the differences 
in the adopted values for the magnetic field strength and maximum proton energy; (ii) the Bethe--Heitler process, which
acts as an injection mechanism of relativistic pairs. Because of the strong magnetic field and large $\gpmx$ used in \cite{Abdo2011}, 
the $\gamma$-ray emission from Mrk~421 is mainly explained as proton and muon synchrotron radiation. In our model, however, these
contributions are not important, and the $\gamma$-rays in the Fermi/LAT and MAGIC energy ranges are explained
mainly by synchrotron emission from $\pg$ pairs (see DPM14). 
Moreover, the emission from hard X-rays up to sub-MeV energies is attributed to different
processes. In our model, it is the result of synchrotron emission from Bethe--Heitler pairs, whereas
in the hadronic model of \cite{Abdo2011}, this is explained as synchrotron emission from pion-induced cascades. It is important to note
that the Bethe--Heitler process was not included in the hadronic model of \cite{Abdo2011}.

As far as the neutrino emission is concerned,
we find that the model-predicted neutrino flux is below but close to the 1 $\sigma$ error bar of 
the derived value for the associated neutrino 9 (PR14).
This proximity is noteworthy and suggests that the proposed leptohadronic model for Mrk~421
can be confirmed or disputed in the near future, as  IceCube collects more data. 
Even in the case of a future rejection of the proposed model for Mrk~421, there 
will still be room
for leptohadronic models that operate in a different regime of the parameter space. For example, in the LHs model,
the neutrino spectrum is expected to 
peak at higher energies and to be less luminous than the one derived here, because  
of the higher values of $\gpmx$ and $B$ (see also in DPM14); this would also hold for the hadronic
scenario used in \cite{Abdo2011}, although no discussion of neutrinos is presented therein. For such
leptohadronic variants, $\sim$PeV neutrino observations cannot be model constraining.

\subsubsection{PG~1553+113}
\label{pg1553}
\begin{figure*}
 \centering
\includegraphics[width=0.6\textwidth]{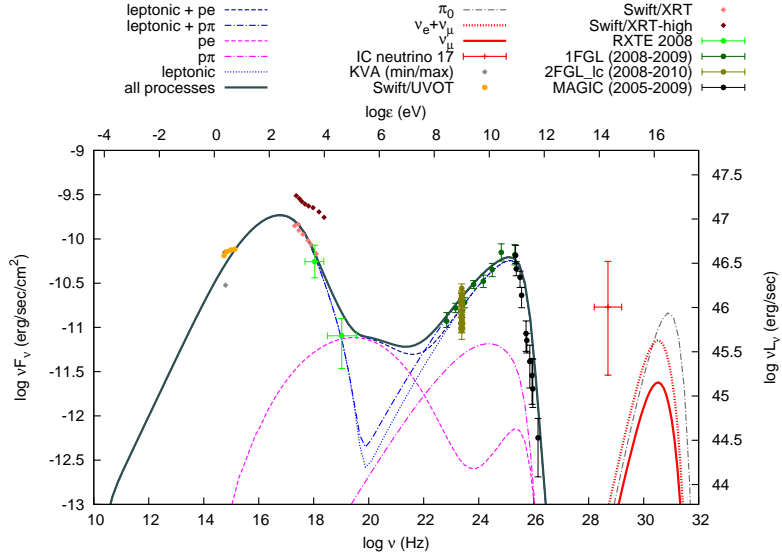}
 \caption{SED of blazar PG~1553+113 for the period 2005-2009 and the neutrino flux for the corresponding IceCube event (ID 17).
 The data are obtained from \citealt{aleksic12}.
 Grey diamonds correspond to the KVA minimum and maximum 
 fluxes and orange circles are the Swift/UVOT observations. 
 Light and dark red diamonds are observations from Swift/XRT in 2005 corresponding 
 to an intermediate and high flux level, respectively. With light green circles
(from top to bottom) are plotted the average RXTE flux and the average 15-150 keV Swift/BAT flux. Green circles represent the average Fermi/LAT spectrum
(August 2008- February 2009) and black circles the MAGIC observations over the periods 2005-2006 and 2007-2009.}
% The grey arrow denotes the 
%  upper limit in 20~keV-40~keV X-rays from the fourth IBIS catalog (\citealt{ubertinietal09}).}
 \label{pg1553+113}
\end{figure*}

The hybrid SED of blazar PG~1553+113 is shown in Fig.~\ref{pg1553+113}. The contribution of 
different emission processes to the total SED is depicted with different types of lines, which
are explained in the legend above the figure. The same applies to all figures that follow. We did
not 
do the same for the case of Mrk~421, since we quoted results originally presented elsewhere, where a detailed description
of the various processes contributing to the overall SED can be found. In any case, it was the prediction
of the model about the neutrino flux from Mrk~421 we wanted to emphasize and not the spectral modelling of the SED.
% }\footnote{
% Our aim in the case of Mrk~421 is to emphasize the prediction of the model about the neutrino flux,
% as originally presented in DPM14. Plotting of all the underlying emitting components would be disorientating, and
% thus we prefer to show only the total photon emission (Fig.~\ref{mrk421}). In any case, can be found in \cite{mastetal13} and DPM14.}
A few things about the hybrid SED are worth mentioning. 
It is the SSC emission of primary electrons (blue dotted line) that mainly contributes to the GeV-TeV $\gamma$-ray regime, whereas
 the $\pg$ component is hidden below it (magenta dash-dotted line). Notice also that the other component of photohadronic origin,
 namely the synchrotron emission from Bethe--Heitler pairs (magenta dashed line), has approximately the same peak flux as the $\pg$ component.
 The fact that the SSC emission is favoured with respect to the emission from photohadronic processes
 has to do with the particular choice of parameter values (Table~\ref{tab-0}). Besides the weak 
 magnetic field ($B=0.05$~G), the combination of 
 a large source ($\rb =2\times10^{17}$~cm) and a relatively high Doppler factor ($\delta = 30$), 
 decreases the compactness of the primary synchrotron photon field, which
 is the target for the photohadronic interactions, and eventually leads 
  to a suppression of the   proton cooling due to photohadronic processes and of the respective emission signatures
 (for more details, see Sect.~3.3 in  PM15).

 In our framework, the total neutrino luminosity is expected to be of the same order of magnitude as
 the synchrotron luminosity of $\pg$ pairs that is emitted as high-energy $\gamma$-rays (see also PM15).
 This is clearly shown in Fig.~\ref{pg1553+113}, where the  peak luminosity of the $\pg$ component
 is similar to the one of the neutrino component  (red dotted line).  Taking into account that 
 the $\pg$ component has a small contribution to the $\gamma$-ray emission, we expect also
 the peak neutrino flux to be lower than the peak $\gamma$-ray flux, which is verified numerically (see Fig.~\ref{pg1553+113}).
 
The previous results may be quantified through the ratio
  $\ygn$, which is defined as 
  \eqb
  \ygn = \frac{L_{\nu}}{L_{\gamma, \rm TeV}},
  \label{Ygn}
  \eqe
  where $L_{\nu}$ and $L_{\gamma, \rm TeV}$ are the total neutrino and  $0.01-1$~TeV $\gamma$-ray luminosities, respectively, as measured
  in the observer's frame. The ratio $\ygn$ 
  expresses the link between the observed $\gamma$-ray luminosity and the model-predicted neutrino luminosity from a blazar, and as such
  is as an important parameter of this study; this will become clear later in Sect.~\ref{ratio}.
  
For PG~1553+113, in particular, we find $\ygn \sim 0.1$, which 
  suggests that the $\pg$ component is not the sole contributor to the observed
 $\gamma$-ray flux.
 Since the low value of $\ygn$ is a matter of parameter values, it might make the reader wonder why we adopted
 the specific parameter values in the first place, and why we did not search for parameter sets that would favour
 photopion production. There are two profound reasons for our particular choice:  the shape of the Fermi/LAT
 spectrum, and the hard X-ray observations with Swift/BAT. Although a
 higher proton injection luminosity or a more compact emitting region would lead to higher neutrino fluxes and could, thus,
 account for the neutrino event ID 17, it would inevitably raise the Bethe--Heitler component. Since the emission from pe
 pairs falls, in general, in the hard X-ray and soft gamma-ray energy bands, the Bethe--Heitler component would not only exceed
 the BAT observations but also destroy the fit to the low-energy Fermi/LAT data. One must also take into account that the 
$\pg$ synchrotron spectrum  is  broader around its peak than the SSC one, which would make it difficult to explain
the observed Fermi/LAT spectrum. 
  
Inspection of Table~\ref{tab-0} reveals that PG~1553+113 is the second 
most energetically demanding 
case, with $L'_{\rm p, inj}$ being approximately one order of magnitude higher than the other values.
In terms of parameter values, such as the size $\rb$ and the magnetic field, it is an ``outlier'' of the sample.
This should not come as a surprise, since PG~1553+113 is the most distant BL~Lac of 
our sample with an estimated redshift $z \sim 0.4$ (see Sect.~\ref{case}), and
its $\gamma$-ray flux is approximately equal to that of Mrk~421.
PG~1553+113 is, therefore, an intrinsically more luminous source than Mrk~421 and requires
an accordingly higher proton injection luminosity as measured in the comoving frame,
unless a larger value for Doppler factor ($\delta \gtrsim 30$) was adopted. For a given $\rb$
this would lower the required proton compactness, and thus
$L'_{\rm p,  inj}$. 
We attempted, however,  to avoid  higher Doppler factors than those listed in Table~\ref{tab-0}, since those
would shift the neutrino peak to hundreds of PeV in energy (see equation~(\ref{Ev})).

Even for the adopted parameters, the peak energy of the model-derived neutrino spectrum
is $\sim 10$~PeV. If we combine this  
with the hardness of the spectrum we obtain a much lower flux than the observed one,
at the energy of the associated neutrino event;
still, this lies within the 3 $\sigma$ error bars (see also Sect.~\ref{neutrino-emission}).

In Fig.~\ref{pg1553+113} we have also included the X-ray high flux state of 2005 for illustrating reasons. An association
of a higher X-ray flux with a higher neutrino flux would be intriguing,  since it would bring
the model-derived flux shown in Fig.~\ref{pg1553+113} even closer to the observed one.
However, such an association is not trivial in the absence of simultaneous observations in high-energies; neither GeV nor
 TeV data were available during the 2005 X-ray high state (see Fig.~1 in \citealt{aleksic12}). Consider for example
 a scenario where the increase in the X-ray flux is caused by an increase of the electron injection
 compactness. This would also result in a higher $\gamma$-ray flux, since
the emission from both the SSC and $\pg$ components would increase. Finally, 
it would lead to a higher neutrino flux because 
of the higher synchrotron photon compactness. However, if the X-ray high flux state was not accompanied by an increase 
in the $\gamma$-ray emission, 
the aforementioned scenario would not be viable. 
More than one model parameters should be changed, and,  depending on their combination, the derived 
neutrino flux would also differ.
Because of the wide range of possibilities we do not, therefore, 
attempt to fit the high-state observed in the X-rays, nor to calculate the respective neutrino
flux.
\begin{figure*}
 \centering
\includegraphics[width=0.6\textwidth]{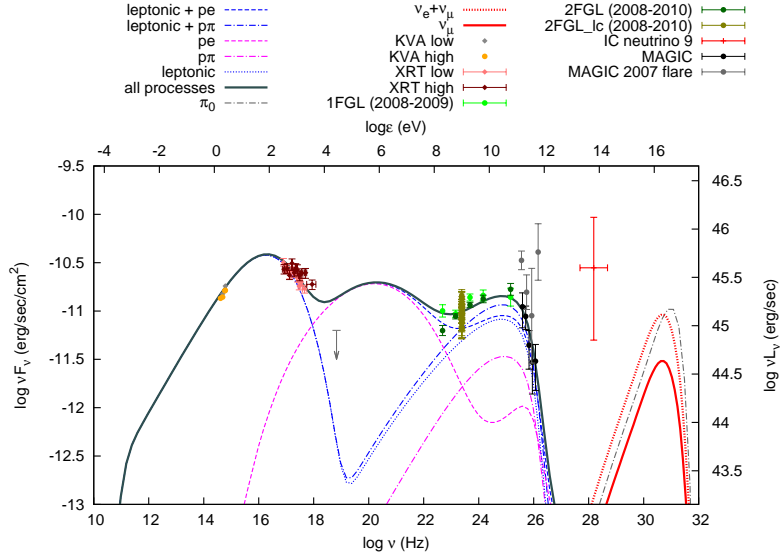}
 \caption{SED of blazar 1ES 1011+496 for the period 2008-2010 and the neutrino flux for the corresponding IceCube event (ID 9).
 Optical, soft X-ray and TeV data were obtained during the first MW campaign in 2008 (\citealt{reinthal12}),  whereas
 GeV observations are from the first (1FGL) and second (2FGL) Fermi/LAT catalogs (quasi-simultaneous). 
 The first TeV detection is over-plotted with grey points
 for comparison reasons, as it corresponds to a high flux state; the respective data points were obtained from ASDC.
 The grey arrow denotes the  upper limit to the 20~keV-40~keV X-rays from the fourth INTEGRAL/IBIS catologue (\citealt{ubertinietal09}).
 }
 \label{1es1011}
\end{figure*}
\begin{figure*}
 \centering
\includegraphics[width=0.6\textwidth]{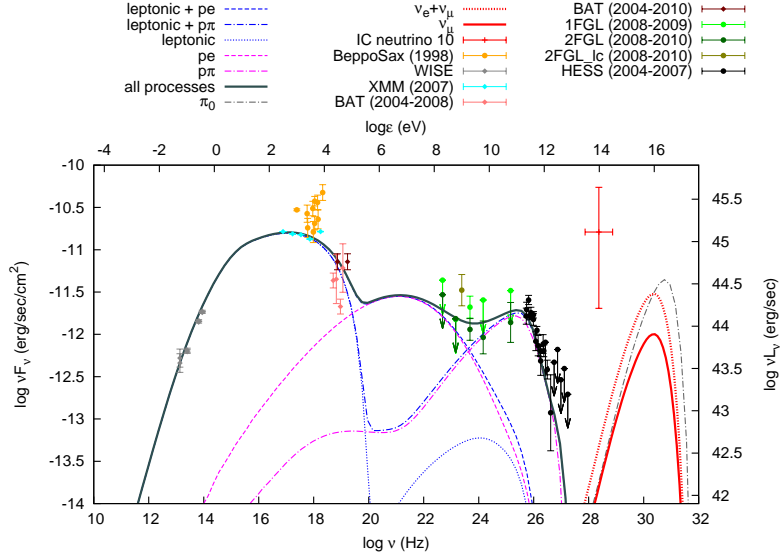}
 \caption{SED of blazar H~2356-309 for the period 2004-2010 and the neutrino flux for the corresponding IceCube event (ID 10).
  The data are obtained from ASI Science Data Center
  using the SED builder tool. All types of symbols and lines are explained in the legend
  above the plot.}
 \label{h2356-309}
 \end{figure*}
 \subsubsection{1ES 1011+496}
\label{1es}
The hybrid SED is shown in Fig.~\ref{1es1011} and the resemblance to the model SED of 
PG~1553+113 is obvious. The $\gamma$-ray emission
is mainly attributed to the SSC emission of primary electrons, since the flux of the $\pg$ component is only a small fraction
of the observed $\gamma$-ray one. The fact that the model neutrino peak
flux is lower than the peak $\gamma$-ray flux should come as no surprise  after the discussion in Sect.~\ref{pg1553}.
The model in Fig.~\ref{1es1011} is mainly constrained by the simultaneous optical, soft X-ray and TeV data of the 2008 MW campaign, since
there are no significant spectral variations in the Fermi/LAT observations covering the period 2008-2010. 
As already noted in PM15, any information on the hard X-ray emission, e.g.
$\epsilon \gtrsim 20$~keV, would be even more constraining for the model. 
This becomes also evident from the cases of PG 1553+113  and H~2356-309 shown in Figs.~\ref{pg1553+113}
and \ref{h2356-309}, respectively.

We searched, therefore, {\sl a posteriori} for possible detections or upper limits
in the hard X-ray band, even if these were not, strictly speaking, simultaneous with the 2008 data. For this, we used
the fourth INTEGRAL/IBIS catalog in 20-40 keV \citep{ubertinietal09},  that covers the period 2003-2008. The source has never been detected 
within this period, which resulted in the upper limit of $\sim 6.1\times 10^{-12}$~erg cm$^{-2}$ s$^{-1}$ shown with a grey arrow in Fig.~\ref{1es1011}. 
The model SED exceeds by a factor of $\sim 3$ the upper limit. This suggests that the photohadronic emission (both pe and $\pg$) should
decrease, if we were to include the upper limit into our fitting procedure. 
Since the most straightforward way to decrease the photohadronic emission in order to accommodate 
the hard X-ray constraint is to assume a lower $\lpinj$, the derived value for $L_{\rm p}^{' \rm inj}$ listed in Table~\ref{tab-0}
should be considered only as upper limit. A lower $\lpinj$ (keeping all other parameters unaltered) would also result in an accordingly lower
neutrino flux and, thus, lower $\ygn$.
Given that the model neutrino flux at the same energy bin with neutrino event 9 
is already below the observed one 
by approximately two orders of magnitude (see Figs.~\ref{1es1011} and \ref{neutrinos}), we argue that
the hard X-ray constraint makes the connection of 1ES~1011+496 with
the neutrino event 9 even less plausible.

 \subsubsection{H~2356-309}
 H~2356-309 is one of the sources in the sample that has also been detected in TeV energies by H.E.S.S.
 and is the most probable counterpart for the IceCube neutrino with ID 10.
 Our model SED and neutrino flux are shown in Fig.~\ref{h2356-309}.
%  , where different
%  types of lines are used to denote the contribution of various emitting
%  components. 
 In contrast to PG~1553+113, the Fermi/LAT spectrum can be approximated as $\nu F_{\nu} \propto \nu^0$ within the error bars and 
 the  $\gamma$-ray flux at $\sim 0.1$~TeV is $\sim 1$ order of magnitude below the peak flux of the low-energy hump. The properties
 of the observed SED of H~2356-309 allows us, therefore, to search for parameters that favour the $\pg$ emission and suppress
 the contribution of the SSC component, contrary to the cases of PG~1553+113 and 1ES~1011+496.
 
By adopting $\rb =3\times 10^{15}$~cm, which is smaller by two and one orders of magnitude than 
 the source's radius used for PG~1553+113 and 1ES~1011+496, respectively, we increase
 the photon compactness of the source. Moreover, 
 the choice of a  stronger magnetic field ($B=5$~G) results in higher magnetic compactness ($\ell_{\rm B}$) than before, where
 $\ell_{\rm B} \propto B^2 / \rb$ and is a measure of the synchrotron cooling rate.
  Thus, the synchrotron emission by secondary pairs
 from pe and $\pg$ processes is enhanced compared to the previous two cases. 
 Figure~\ref{h2356-309} displays the above. It is, indeed, the emission from the $\pg$ pairs that mainly contributes
 to the observed $\gamma$-ray flux, while the SSC component is $\sim 1.5$ orders of magnitude less
 luminous. Moreover, the ``Bethe--Heitler bump'' becomes a prominent feature of the SED. Both
  results suggest a higher efficiency of both photohadronic processes compared to the previous two cases.

The neutrino spectral shape is similar to the previous cases and peaks at $\sim 10$~PeV.
However, the ratio $\ygn$ increases  to $0.8$, 
 an eightfold increase from the case of PG~1553+113, which reflects the different efficiencies of photopion production in these cases.
 Although $\ygn$ is close to unity, 
 the peak neutrino peak flux is higher than the peak $\gamma$-ray flux because
 of the different spectral shapes of the two components. At the energy of the detected neutrino  ($\sim 0.1$~PeV)
 the model-derived flux is low, but lies within the 3~$\sigma$ error bars. This discrepancy is similar
 to the one we found in previous cases, such as PG~1553+113, although the photopion production efficiency and $\ygn$
are both higher in the case of H~2356-309. 
At first sight, this result seems contradictory. 
However, one has to take also into consideration the differences in the 
 ratios of the observed $\gamma$-ray and neutrino fluxes. For example, in blazars PG~1553+113 and H~2356-309, 
 this ratio is close to and much less than unity, respectively. Thus, in the case of H~2356-309, the higher $\ygn$
 is compensated by the lower ratio of observed $\gamma$-ray and neutrino luminosities.
  
\begin{figure*}
 \centering
 \includegraphics[width=0.6\textwidth]{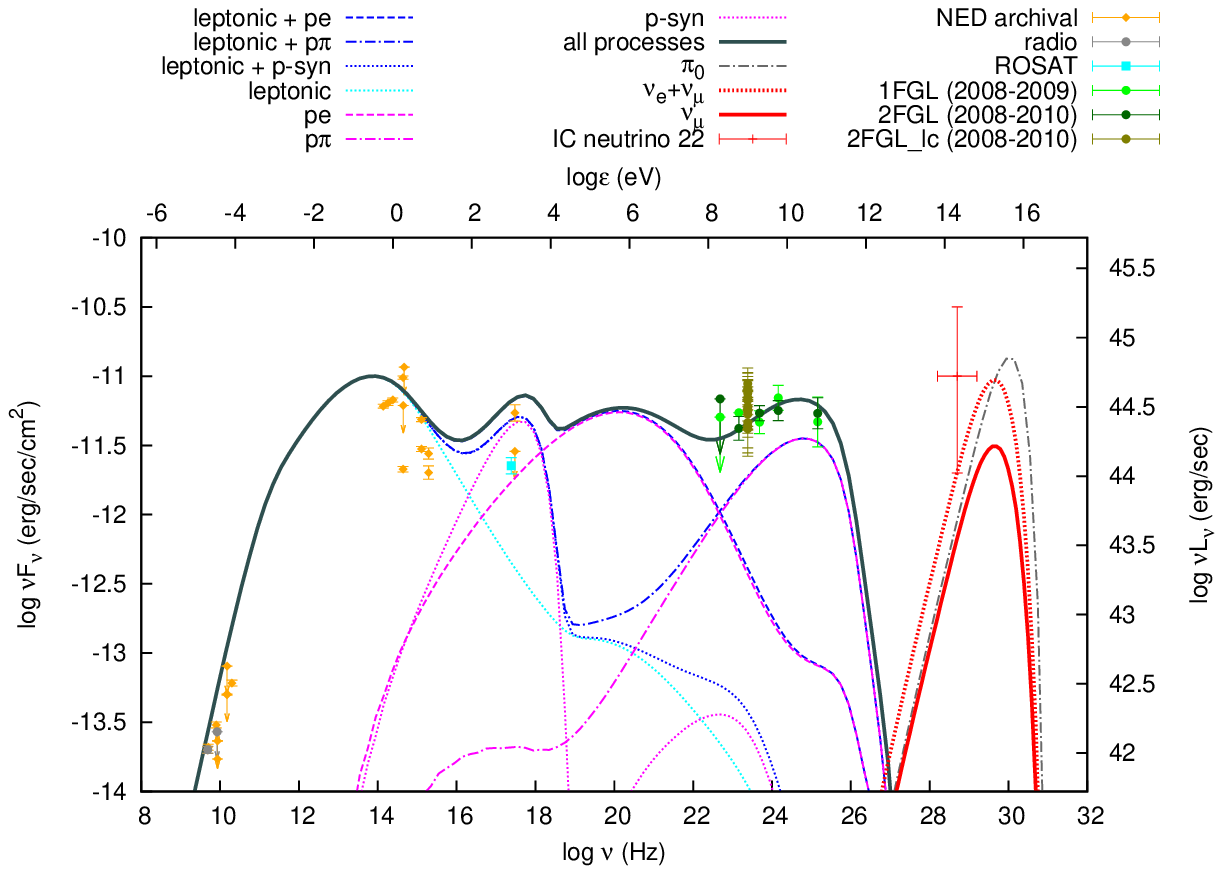}
 \caption{SED of blazar H~1914-194 and the neutrino flux for the corresponding IceCube event (ID 22).
 The data are obtained from ASI Science Data Center
  using the SED builder tool. All types of symbols and lines are explained in the legend
  above the plot.}
 \label{h1914-194}
\end{figure*}
\begin{figure*}
 \centering
\includegraphics[width=0.6\textwidth]{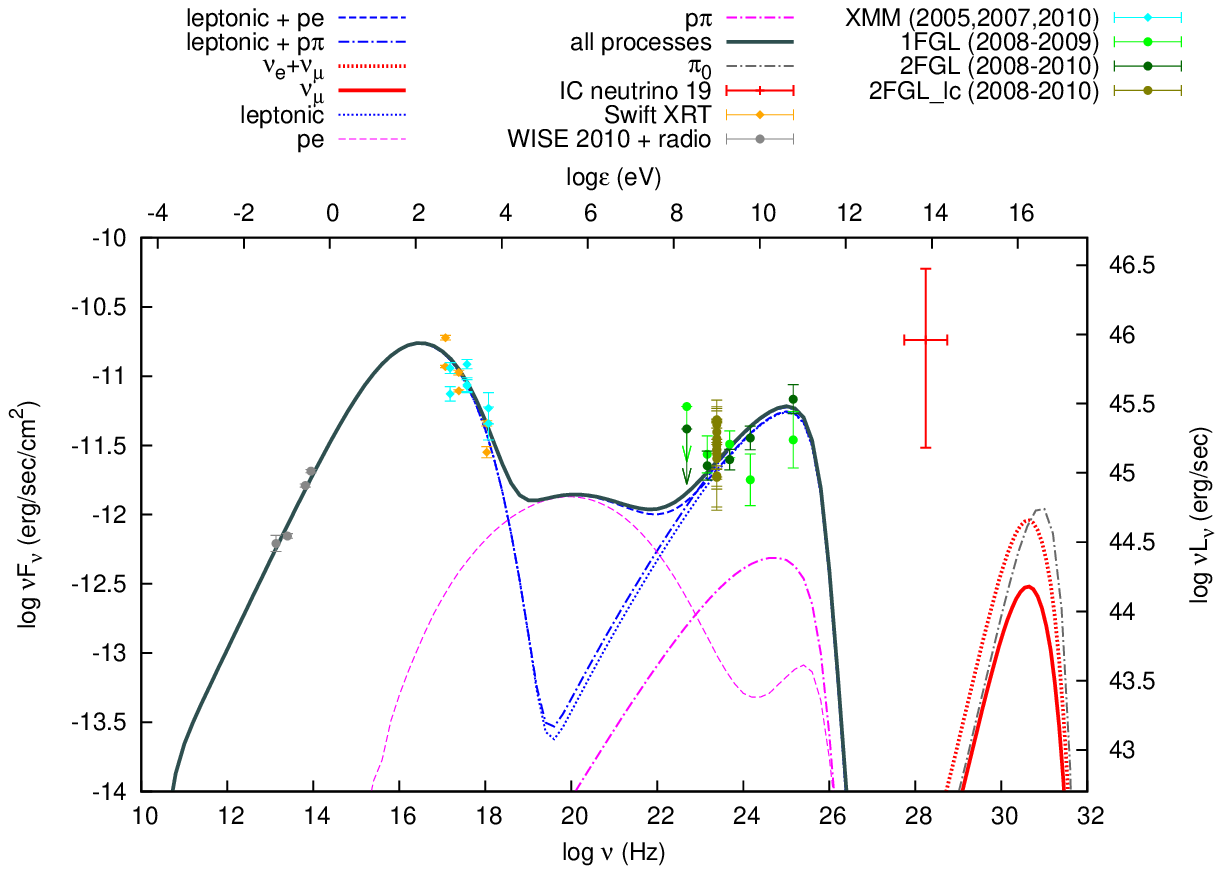}
 \caption{SED of blazar 1 RXS~J054357.3-553206 and neutrino flux for the corresponding IceCube event (ID 19).
The data are obtained from ASI Science Data Center
  using the SED builder tool. All types of symbols and lines are explained in the legend
  above the plot.}
 \label{rxs}
\end{figure*}
 \subsubsection{1H~1914-194}
 \label{h1914}
Blazar 1H~1914-194 is one of the two sources that are potential TeV emitters  and
has the poorest MW coverage among the BL~Lacs of the sample (for details, see Sect.~\ref{case}).
Because of the inhomogeneity (in time domain) of the available MW observations, 
we present only
an indicative fit, where we attempted to describe at best the Fermi/LAT data.
Being the least constrained case we chose parameters that ensure high 
efficiency in photopion production, i.e. we present
the most optimistic, for neutrino emission, scenario.
We caution the reader that contemporaneous data in optical, X-rays and GeV $\gamma$-rays
may strongly constrain the model (see e.g. Mrk~421 and PG~1553+113)
and lead to different parameter values
from those listed in Table~\ref{tab-0}. The neutrino flux shown in Fig.~\ref{h1914-194}
should be therefore considered  as an upper limit. We also 
note that the non-detection of TeV $\gamma$-rays is not as important for the predicted
neutrino flux as the simultaneous coverage in the aforementioned energy bands. For example,
the peak neutrino energy is solely determined by the peak frequency
of the low energy component and the value of the Doppler factor (see equation (\ref{Ev})), while
the neutrino flux is related to the sub-TeV $\gamma$-ray flux (see Figs.~\ref{pg1553+113}, \ref{h2356-309} and Fig.~8 in PM15).

The SED we obtained is the most complex one in terms of the number of emission
components that appear in the MW spectrum.
Apart from the very low frequencies (below UV) where the primary leptonic component
contributes, the spectrum is comprised roughly of: proton synchrotron radiation (between UV and $\sim$keV),
synchrotron emission from Bethe--Heitler pairs (between $\sim$keV and $\sim$100 MeV) and synchrotron emission  ($\lesssim 1$~TeV) from
secondary pairs produced through pion decay and $\gamma \gamma$ absorption of VHE $\pi^0$ $\gamma$-rays (see equation (14) 
in \citealt{petromast12}).
In fact the role of $\gamma \gamma$ absorption is non-negligible in this case, since the unattenuated
$\pi^0$ component has a flux higher by a factor of 
$\sim 2$ than the neutrino one (see dash-dotted grey curve). The attenuation of 
VHE $\gamma$-rays initiates an EM cascade \citep{mannheim91} that
transfers energy from the PeV regime to lower energies. Our numerical approach does not 
allow us to isolate multiple photon generations produced 
in the electromagnetic cascade. Thus, strictly speaking, the pe and $\pg$ components shown in Fig.~\ref{h1914-194} 
are the result of synchrotron radiation not only from pairs produced through photohadronic interactions but
also from pairs produced in the EM cascade.

In this case, we find $\ygn \simeq 2$, 
which is the highest value amongst the sample (see Table~\ref{tab-0}) and close
to the maximum value ($\ygn^{\max}=3$) that is allowed in this theoretical framework (see Sect. 3.2 in PM15).
The model-predicted neutrino flux is close to the observed flux (within the 1 $\sigma$ error bars), thus strengthening 
the association of neutrino event 22 with this source. We have to note, however,
that the model-predicted neutrino flux might decrease, should more constraints on the SED be available in the future.

\subsubsection{1 RXS~J054357.3-553206} 
\rxs \ is the second HBL source that has not yet been detected in TeV energies and the most probable
counterpart of IceCube event 19 according to PR14. 
The hybrid SED obtained with our model is shown in Fig.~\ref{rxs} and shares many features
with the model SED of PG~1553+113. 
According to the discussion in Sect.~\ref{pg1553}, we can infer
that the observed $\gamma$-ray emission originates only partially
from photohadronic processes, since the model derived neutrino peak flux
is lower than the $\gamma$-ray one. In fact, we find that $\ygn=0.1$, same as in the case
of PG~1553+113. However, because the flux corresponding to neutrino event ID 19 is higher than the $\gamma$-ray one, the
model-predicted flux (at the same energy bin as event 19) is more than 3 $\sigma$ off from the observed value, and 
the model prediction falls short of explaining the observed flux (see also Fig.~\ref{neutrinos}).

 In contrast to PG~1553+113 the SED of blazar \rxs \ is less constrained both in (sub)MeV and TeV energies. 
This, along with the large discrepancy between the model and observed neutrino fluxes, makes \rxs \ the best
case for studying the effects of a different parameter set on the neutrino flux. 
We search, therefore, for parameters that increase the contribution of the $\pg$ component to the observed
$\gamma$-ray emission and at the same time lead to a higher neutrino flux. We do not focus, however, on the goodness of the SED fit but we rather
try to roughly describe the observed spectrum.
The aim of what follows is (i) to demonstrate the flexibility of the leptohadronic model in describing the MW photon spectra and (ii)
to find parameters that maximize the predicted neutrino flux from the particular blazar, albeit 
at the cost of a worse description of the SED.

\begin{figure}
 \centering
\includegraphics[width=0.48\textwidth]{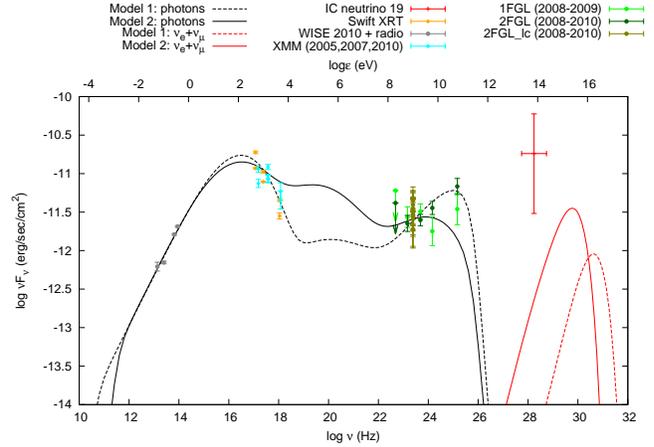}
 \caption{SED of blazar 1 RXS~J054357.3-553206 and neutrino flux for the corresponding IceCube event (ID 19).
The hybrid model SEDs obtained for two different parameter sets presented in Tables~\ref{tab-0} and \ref{tab-1} are respectively shown
with dashed and solid lines.
}
 \label{rxs-comp}
\end{figure}

Our results are summarized in Fig.~\ref{rxs-comp}, where we plot the
hybrid SED obtained for a different parameter set presented in Table~\ref{tab-1}.
From this point on, we will refer to it as Model~2.
For a better comparison, we over-plotted with dashed lines 
the photon and neutrino spectra that were obtained for the parameters listed in Table~\ref{tab-0} (Model~1).
In Model~2 the $\gamma$-ray emission is mainly attributed to the $\pg$ component and the neutrino flux is
comparable to the observed $\gamma$-ray one; in particular, we find $\ygn \simeq 3$.
In our framework, this is the most optimistic scenario and the derived neutrino flux can be considered as an upper limit on the flux
expected from this source (see also discussion in Sect.~\ref{h1914}).
However, the description of the SED is not as good as in our baseline model. In particular, the model-derived spectrum 
(i) cannot explain the highest energy
Fermi/LAT data,  (ii) is less hard (in $\nu F_{\nu}$ units) at GeV energies  and more broad 
than the observed one -- see also Sect.~\ref{pg1553}, and  (iii)  is marginally above the highest flux observation in X-rays.
This excess is caused by the
synchrotron emission from Bethe--Heitler pairs, which appears in 
the hard X-ray/sub-MeV energy range, and it is even more luminous than the $\pg$ component.
It is also noteworthy that the Bethe--Heitler emission in Model~2 plays a crucial role in 
the neutrino production: if we were to neglect pair injection from the Bethe--Heitler process,
the neutrino flux in Model~2 would be lower by a factor of $\sim 10$.
The detection of a luminous component in the 20~keV--20~MeV energy range
could, therefore, serve as an indirect probe of high-energy neutrino emission from a 
blazar source within the leptohadronic scenario, as recently proposed in PM15.

 \subsection{Neutrino emission}
 \label{neutrino-emission}
 %%% In figure neutrinos_c.eps I plot the muon+electron fluxes -- in previous version some curves
 %%% were showing by mistake the muon only!!!
 \begin{figure*}
 \centering
 \includegraphics[width=0.6\textwidth, height=8cm]{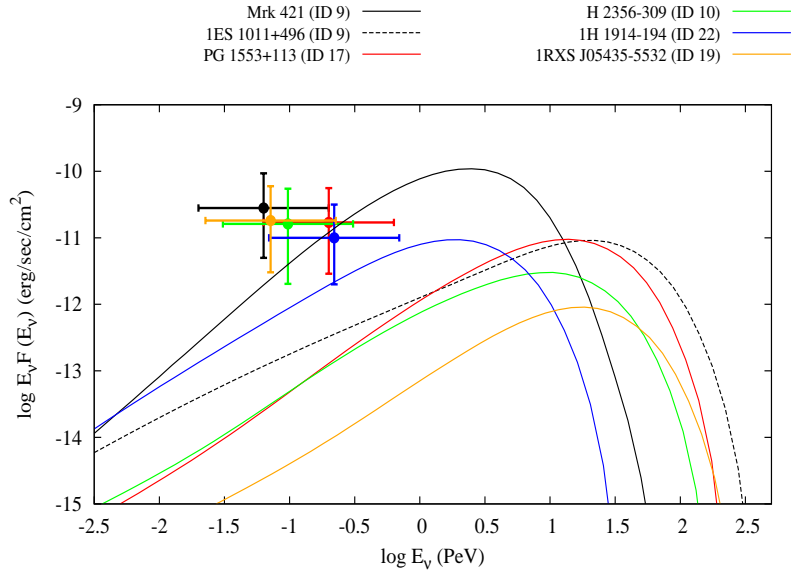}
 \caption{Comparison of the model (lines) and the observed (circles) neutrino fluxes as defined in PR14 for 
 the six BL~Lacs of the sample. 
 %Symbols with the same color correspond to the same source. 
 The Poissonian 1 $\sigma$ error bars for each event, which are adopted by PR14, are also shown. We note that the respective 3 $\sigma$ error bars 
 correspond to -2.9, +0.9 dex \citep{gehrels86}.}  
% For a better comparison, open squares show the neutrino flux of the model at the same energy bin
%  as the detected neutrino events.}
 \label{neutrinos}
 \end{figure*}
 
 The model-derived electron and muon $\nu+\bar{\nu}$ spectra for 
 the six blazars of the sample are summarized in Fig.~\ref{neutrinos}. For a better comparison, we over-plotted
 the fluxes that correspond to the respective neutrino events along with the Poissonian $1 \sigma$ error bars calculated for a single event
 (PR14). We note that respective 3 $\sigma$ error bars  correspond to -2.9, +0.9 dex \citep{gehrels86}.
 
 We focus first on the neutrino event 9, which according to the model-independent analysis of PR14, has two  plausible astrophysical counterparts;
 namely
 blazars Mrk~421 and 1ES~1011+496. 
 For Mrk~421, in particular, we show the neutrino spectrum depicted in the right panel of Fig.~\ref{mrk421}.
 According to the discussion in Sect.~\ref{1es1011},
 the neutrino spectrum for 1ES~1011+496 (dashed line) should be considered as an upper limit,
 since inclusion of additional constraints in the SED fitting would
 lower the neutrino flux by at least a factor of $\sim3$. Thus, 
 our results strongly favour Mrk~421 against 1ES~1011+496.
 As already discussed in the respective paragraphs of Sect.~\ref{results}, the differences between the neutrino fluxes
 originate from the differences in their SEDs. Thus, the case of neutrino ID 9 reveals in the best way how
 detailed information from the photon emission may be used to lift possible degeneracies between multiple astrophysical counterparts.
  
 \rxs \ was selected by PR14 as the astrophysical counterpart of neutrino event 19. Figure \ref{neutrinos}
 shows the respective neutrino spectrum (yellow line) we  obtained for the best-fit model of its SED. The model-derived  flux
 at the energy of $\sim 0.2$~PeV lies below the 3 $\sigma$ error bars. 
 In this regard,  \rxs  \ may be excluded at the present time from being the astrophysical counterpart
of its associated neutrino event, namely event  19. We note, however, that a
 higher neutrino flux (within the 3 $\sigma$ error bars)
 can be obtained at the cost of a worse description of the SED (see Sect.~\ref{rxs}).
 
 In all other cases, the model-derived\footnote{We refer to the neutrino emission calculated 
 for best-fit SED models.} neutrino flux at the energy bin of the detected neutrino
 is  below the 1 $\sigma$ error bars, but still within the 3 $\sigma$ error bars. 
 Thus, strictly speaking, the association of these sources cannot be excluded at the present time. 
In particular, Mrk~421 and 1H~1914-194 are the two cases
 with the smallest discrepancy between the observed and model-derived fluxes and, in this regard, 
 the most interesting, because their association with the respective IceCube events
 can be either verified or disputed in the near future. 

Besides the comparison between the model and the observed neutrino fluxes, Fig.~\ref{neutrinos} also 
demonstrates the similarity of the model neutrino spectra, despite the fact that they were
obtained for BL~Lacs with different properties. 
Our results suggest that the shape of the neutrino spectrum depends only weakly on 
the $\gamma$-ray luminosity of the source, which varies approximately two orders
of magnitude among the members of the sample (see Table~\ref{tab-0}).
In particular, the neutrino spectrum may 
be described as a power-law, i.e.
$E_{\nu}^2 dN_{\nu}/dE_{\nu} \propto E_{\nu}^{\alpha}$ with $\alpha\sim 1.1-1.3$.

Before closing this section, we must point out that 
the present theoretical framework allows us to establish a 
connection between the {\sl observed} $\gamma$-ray luminosity and the {\sl predicted} total neutrino luminosity.
In fact, we showed that these are simply proportional as $L_{\nu} = \ygn  L_{\gamma, \rm TeV}$, with 
 $\ygn$ in the range $0.1-2$ (Table~\ref{tab-0}). Although 
 there is a maximum value predicted by the model, i.e. $\ygn \lesssim 3$ (see also PM15),
 there is no restriction on the lowest possible value of this ratio. 
 Values of $\ygn \ll 1$ will only imply that the leptohadronic model simplifies into
 a  pure leptonic one, with the SSC emission of primary electrons being responsible for the observed
 emission. This is illustrated best through the cases of  PG~1553+113, 1ES~1011+496, and \rxs.

\begin{table}
\centering
 \caption{Parameter values used for the fit presented in Table~\ref{tab-0}  (Model 1) and in the fiducial model described in text, which maximizes
 the neutrino flux (Model~2) from blazar \rxs. The symbols have the same meaning as in Table~\ref{tab-0}.}
 \begin{tabular}{c c c}
  \hline \hline
  Parameter & Model~1 & Model~2 \\
  \hline \hline
  B & 0.1 & 1 \\
  $\rb$ (cm)& $3\times 10^{16}$ & $3\times 10^{15}$ \\
  $\delta$ & 31 & 50 \\
    \hline
  $\gamma_{\rm e, \min}$ & 1 & 1 \\
  $\gamma_{\rm e, \max}$ & $1.2 \times 10^5$ & $2.5\times 10^4$ \\
  $s_{\rm e}$ & 1.7 & 1.7 \\
  $\leinj$ & $2.5 \times 10^{-5}$ & $1.6 \times 10^{-5}$ \\
  %$L'_{\rm e, inj}$~(erg/s) & $1.2 \times 10^{42}$& $7.6\times 10^{40}$\\
  \hline
  $\gamma_{\rm p, \min}$ & 1& 1\\
  $\gamma_{\rm p, \max}$ & $1.2 \times 10^7$ & $1.2\times 10^6$\\
  $s_{\rm p}$ & 2.0 & 1.9 \\
  $\lpinj$ & $8 \times 10^{-4}$& $1.6 \times 10^{-2}$\\
 % $L'_{\rm p, inj}$~(erg/s) & $5 \times 10^{45}$& $1.1 \times 10^{46}$\\
  \hline \hline
  $L_{\gamma, \rm TeV}$~(erg/s)&  $6.3\times 10^{45}$ & $1.6\times 10^{45}$ \\
  $L_{\nu}$~(erg/s) & $6.3\times 10^{44}$ & $5\times 10^{45}$ \\
   $\ygn$  & 0.1 & 3.1 \\
   \hline
 \end{tabular}
\label{tab-1}
\end{table}

\subsection{The ratio $\ygn$ and the energetics}
\label{ratio}

In this section we examine in more detail the 
results presented in  Table~\ref{tab-0}. We attempt to extract information about
the dependence of 
the ratio $\ygn$, which is a crucial derivable quantity of our model, 
on observables, such as the photon index in the Fermi energy range and the $\gamma$-ray luminosity $L_{\gamma, \rm TeV}$.
Moreover, using the parameter values of Table~\ref{tab-0}, we calculate the jet power for each source
and comment on the 
energetics.
The results that follow, particularly those related to 
trends and correlations, should be considered with caution because of the limited size of the sample. Although these cannot
be directly applied to a wider sample of BL~Lacs, they are useful in that they provide
a better understanding of the modelling results.
 \begin{figure}
 \centering
 \includegraphics[width=0.47\textwidth]{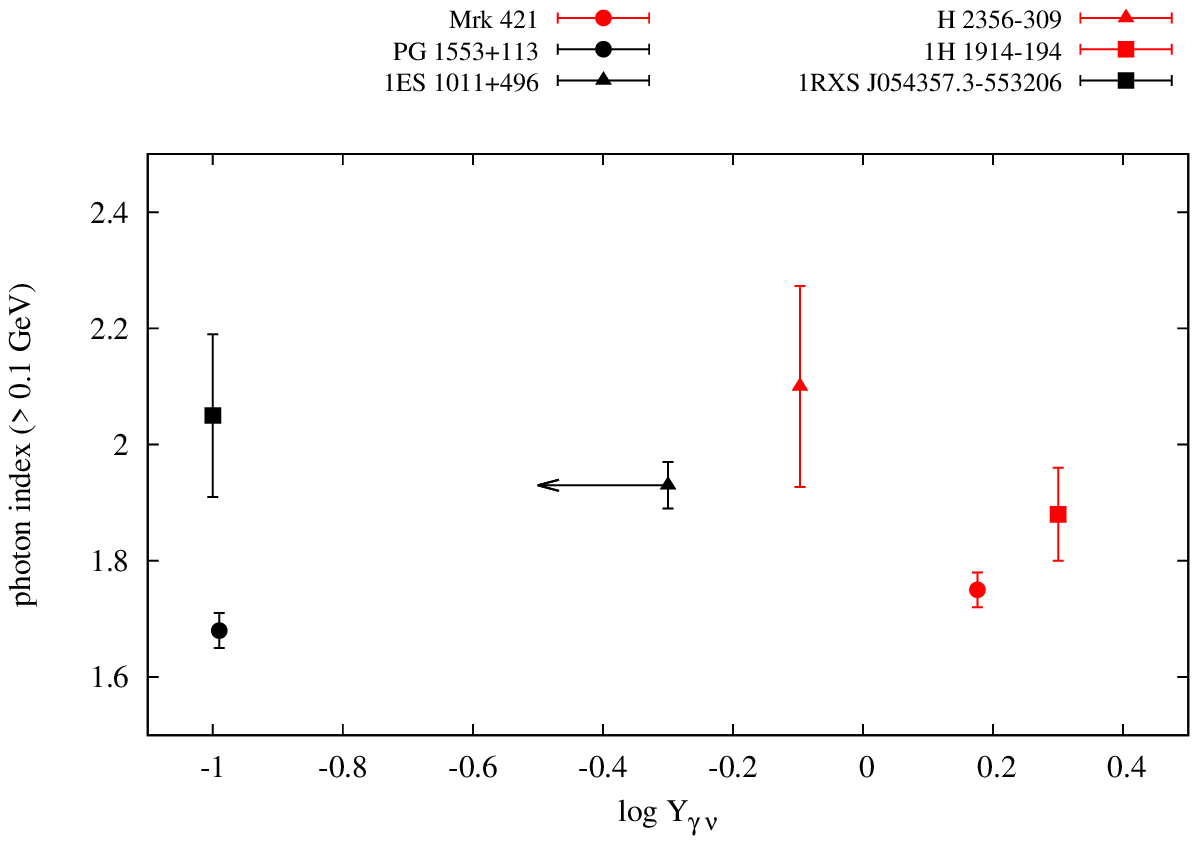} 
  \includegraphics[width=0.48\textwidth]{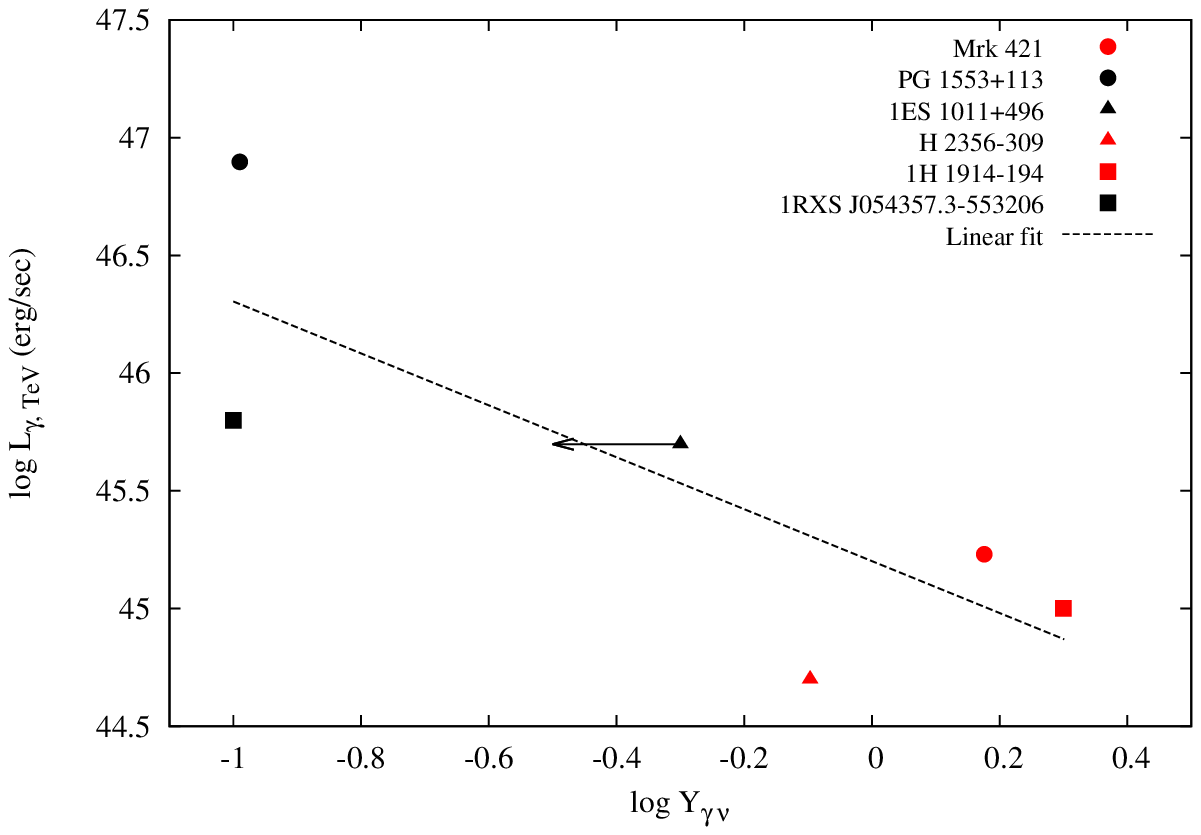} 
 \caption{Top panel: Plot of the photon index in the Fermi/LAT energy range versus the ratio $\ygn$.
 Sources with ratio values above and below the mean value $\bar{Y}_{\gamma \nu}=0.8$ are plotted with red and black symbols, respectively.
 The arrow stresses the fact that the ratio derived for blazar 1ES~1011+496 is only an upper limit (see Sect.~\ref{1es1011}).
 The values of the photon index are adopted from the 1FGL catalog (\citealt{abdo10}). Bottom panel: The observed 
 (0.01-1) TeV  $\gamma$-ray luminosity plotted against the ratio $\ygn$ (logarithmic scale). The arrow and black (red) symbols 
 have the same meaning as 
 in the top panel. The dashed line is a linear fit ($\log L_{\gamma, \rm TeV} = A_1\log \ygn +A_0$) to the points
 with  $A_1=-1.1 \pm 0.4$, $A_0=45.2 \pm 0.2$.}
%  if we use either $\ygn=0.5$ (black line) or $\sim 0.16$ (blue line)  for blazar 1ES 1011+496.  
%  The parameters of the fits are: $A_1=-1.1 \pm 0.4$, $A_0=45.2 \pm 0.2$ (black line) 
%  and  $A_1=-1.03 \pm 0.4$, $A_0=45.1 \pm 0.2$ (blue line).}
 \label{Y_phindex}
 \end{figure}
 
\begin{figure}
 \centering
 \includegraphics[width=0.48\textwidth]{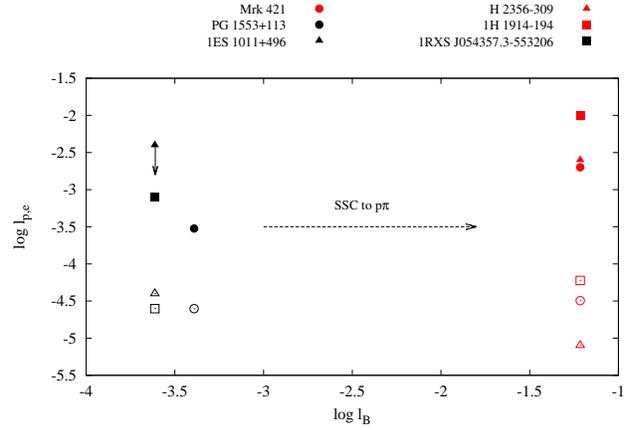}
 \caption{Logarithmic plot of the proton (filled symbols) and electron (open symbols)
 injection compactnesses as a function of the magnetic compactness $\ell_{\rm B}$ (for the definition, see text) for the 
 six BL~Lacs of the sample. Red (black) colored symbols correspond to ratio values $\ygn$ above (below) 
 the average value derived from the sample, i.e. $\bar{Y}_{\nu \gamma} =0.8$. The $\lpinj$ value
 for 1ES~1011+496 is an upper limit, shown with an arrow.
 The dashed arrow demonstrates the transition from
 SSC to $\pg$ dominated $\gamma$-ray emission. }
 \label{lB}
 \end{figure}
  \begin{figure}
 \centering
 \includegraphics[width=0.48\textwidth]{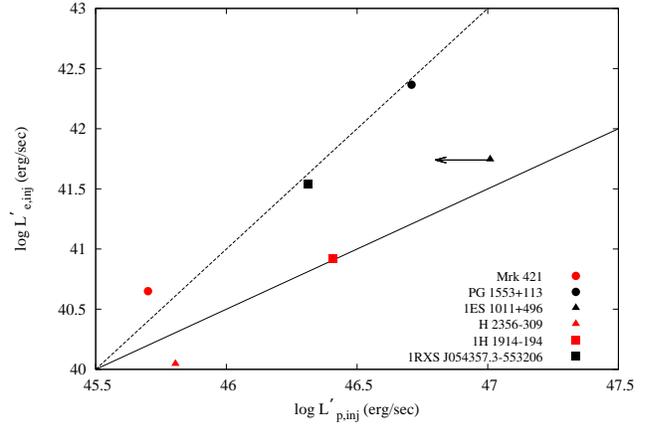}
 \caption{$L'_{\rm e, inj}$ versus $L'_{\rm p, inj}$ in logarithmic scale. The color coding is the same as in Fig.~\ref{lB}.
 The solid and dashed lines have slopes equal to one and two, respectively, and are plotted for guiding the eye.}
 \label{LpLe}
 \end{figure}
 
%    \begin{figure}
%  \centering
%  \includegraphics[width=0.48\textwidth]{./figs/jet.eps}
%  \caption{Jet luminosity (given by eq.~(\ref{jet1})) versus the ratio $\ygn$ for the six blazars of the sample. 
%  The color coding is the same as in Fig.~\ref{lB}. In the case of 1ES~1011+496,  both $L_{\rm j}$ and $\ygn$ 
%  are upper limits; this is illustrated with the arrows.}
%  \label{jet}
%  \end{figure}
 
 In the top panel of Fig.~\ref{Y_phindex} we plot the photon index in the Fermi/LAT energy range 
 \citep{abdo10} versus the ratio $\ygn$. 
 Sources with ratio values
 above and below the mean value of the sample $\bar{Y}_{\nu \gamma}=0.8$ are plotted with black and red symbols, respectively.
  We remind the reader that for 1ES~1011+496 the ratio $\ygn=0.5$ was derived without including in our analysis 
  the upper limit to the hard X-rays. As discussed in Sect.~\ref{1es1011}, inclusion of the hard X-ray constraint would
  lower the current ratio, at least by a factor of $\sim 3$; this is illustrated with an arrow in both panels of Fig.~\ref{Y_phindex}.
  There is no evident trend between the ratio $\ygn$ and the
 photon index in the GeV  energy range, although in Sect.~\ref{pg1553} we used 
 the hardness of the $\gamma$-ray spectrum as an argument for choosing parameters that 
 favoured SSC against $\pg$ radiation. These findings suggest that the Fermi photon index alone
 is not a good indicator of  the model-derived ratio $\ygn$. However, 
 we find a negative trend between
 the observed $\gamma$-ray luminosity in the 0.01-1 TeV energy range (listed in Table~\ref{tab-0})
 and the ratio $\ygn$ (bottom panel in Fig.~\ref{Y_phindex})\footnote{One could argue that
 we were bound to find an anti-correlation, since we are plotting $L_{\gamma, \rm TeV}$ against the ratio $\ygn$
 (note that $L_{\gamma, \rm TeV} = \ygn^{-1} L_{\nu}$). This would be, indeed, correct, if there was an {\sl a priori}
 knowledge of the model-derived neutrino luminosity. In fact, there was no guarantee that the neutrino luminosity obtained
 for all six BL~Lacs would be approximately constant (see Table~\ref{tab-0}), which is an interesting outcome of the present analysis.}.
 The size of our sample is limited, yet
 our results suggest that the contribution of the $\pg$ component to the blazar's $\gamma$-ray emission 
 is smaller in sources that are more $\gamma$-ray luminous. 
 The dependence of $\ygn$ on the $\gamma$-ray luminosity 
 is particularly useful if we attempt to 
 calculate the BL~Lac diffuse neutrino flux by extrapolating
 the findings of this sample to the wider class of BL~Lacs; this will be part of a future study. 
 Finally, we point out that the negative trend shown in Fig.~\ref{Y_phindex} 
 does not necessarily mean that less powerful BL~Lacs in  $\gamma$-rays inject more power in relativistic protons, as it 
 will become evident at the end of this section.
 
The bottom panel of Fig.~\ref{Y_phindex} provides the first hints of a separation between the sources of the sample with ratios 
above and below the average value. However, this becomes clear in Fig.~\ref{lB}, where we plot 
the proton (filled symbols) and electron (open symbols) injection compactnesses
against the magnetic compactness defined as $\ell_{\rm B}= \sth \rb B^2 / 8 \pi \mel c^2$.
Red and black colored symbols are used to denote values of $\ygn$ above and below the mean value, respectively.
Figure~\ref{lB} reveals the presence of two sub-groups within the sample. As $\ell_{\rm B}$ increases,
the synchrotron cooling of both primary and secondary electrons is enhanced.  On the one hand, this 
leads to an increase of the synchrotron emission from pairs produced by $\pg$ interactions that emit in the $\gamma$-ray regime,
while on the other hand, it suppresses the SSC emission of primary electrons. Thus, as $\ell_{\rm B}$ increases, the contribution 
of the $\pg$ component to the $\gamma$-ray emission increases against the SSC one. This is finally reflected in 
an increase of the ratio $\ygn$.
%We cannot, however, draw a definite line between the two sub-groups because of the small sample.

% From Fig.~\ref{lB}  one can deduce that there is an approximate linear relation between
% $\lpinj$ and $\leinj$ used to model the SEDs. This is better illustrated in

In Fig.~\ref{LpLe} we  plot $L_{\rm e}^{'\rm inj}$ as a function of $L_{\rm p}^{'\rm inj}$. 
The color coding is the same as in Fig.~\ref{lB} and the arrow stresses the fact
that the derived value for 1ES 1011+496 is an upper limit (see Sect.~\ref{1es}).
The injection luminosities of primary particles, as measured in the comoving frame, follow a more than 
linear relation, i.e. 
%In any case, the relation between the injection luminosities is 
$L_{\rm e}^{' \rm inj} \propto \left(L_{\rm p}^{'\rm inj} \right)^x$ 
with $1 \le x \le 2$. More important, though, is that
sources with $\ygn \ge 0.8$ lie at the lower left corner of the plot, i.e. they are described
by low proton and electron luminosities (in the comoving frame) with respect to the BL~Lacs with smaller $\ygn$. 

The jet power is a useful quantity when it comes to the energetics of the model. 
In the one-zone leptohadronic model, the energy densities of 
both relativistic electrons ($u'_{\rm e}$) and protons ($u'_{\rm p}$), as well as the energy density of the  
magnetic field ($u'_{\rm B}$) and radiation ($u'_{\rm r}$), contribute to the jet power, which may be written as
\eqb
L_{\rm j} \simeq \pi \rb^2 \delta^2 \beta c \left(u'_{\rm e} + u'_{\rm p} + u'_{\rm B} + u'_{\rm r}\right),
\label{jet1}
\eqe
where the energy densities of the various components are measured at steady state and in the comoving frame, $\delta \approx \Gamma$,
and $\beta = \sqrt{1-1/\Gamma^2} \simeq 1$. For the assumptions used to derive equation~(\ref{jet1}), see e.g. \cite{celottighisellini08}. 
The jet luminosity as 
well as $u'_{\rm i}$ are listed, for reference, in Table~\ref{tab-2}.
For all sources we find
high jet luminosities, and emission regions far from equipartition $u'_{\rm p}/u'_{\rm B} \gg 1$, with the proton component being
the dominant one. These results are typical of leptohadronic models (e.g. \cite{boettcherreimer13, cerrutietal14}) and often constitute a 
point for criticism. 
However, our analysis highlighted that sources with  $\ygn \sim 1$, namely with significant contribution of the photohadronic component to 
the $\gamma$-ray emission, are described by less extreme, i.e. $\lesssim 10^{49}$~erg/s, jet powers and have lower $u'_{\rm p}/u'_{\rm B}$ ratios;
this was not expected beforehand. For the rest of the sources, i.e. those with $\ygn \ll 1$, the derived jet luminosities exceed 
$10^{49}$~erg/s, mainly because of the larger $\rb$ compared to the other sources  (see Table~\ref{tab-2}).

\begin{table*}
\centering
 \caption{Jet luminosity as defined in equation~(\ref{jet1}) and (comoving) energy densities
 of the proton, electron, radiation and magnetic components.}
 \begin{tabular}{l c c c c c}
  \hline \hline
  Source & $L_{j}$ (erg/s)  &  $u'_{\rm p}$ (erg/cm$^3$) & $u'_{\rm e}$ (erg/cm$^3$) & $u'_{\rm r}$ (erg/cm$^3$) & $u'_{\rm B}$ (erg/cm$^3$)\\
  \hline \hline
 Mrk~421 & $2\times 10^{48}$ & $4\times 10^3$  & $4\times 10^{-3}$& $4.9\times 10^{-2}$ & $9.9\times 10^{-1}$ \\
 PG~1553+113 & $3.7 \times 10^{49}$ & $1.1 \times 10$ & $4 \times 10^{-4}$ & $2.3 \times 10^{-4}$ & $9.9 \times 10^{-5}$ \\
 1ES~1011+496 & $8.3 \times 10^{49}$ & $9\times 10^2$ & $6.6 \times 10^{-3}$ & $5.3 \times 10^{-4}$ & $4 \times 10^{-4}$ \\
 H 2356-309 & $4.4 \times 10^{48}$ & $5.7 \times 10^3$ & $1.4 \times 10^{-3}$ & $1.6 \times 10^{-2}$ & $9.9\times 10^{-1}$ \\
 1H 1914-194 & $8 \times 10^{48}$ & $2.9 \times 10^4$ & $7.4 \times 10^{-2}$ & $8.2 \times 10^{-2}$ & $9.9\times 10^{-1}$\\
 \rxs & $1.5 \times 10^{49}$ & $1.8 \times 10^2$ & $3.3 \times 10^{-3}$ & $7\times 10^{-4}$ &  $4 \times 10^{-4}$ \\
   \hline
 \end{tabular}
\label{tab-2}
\end{table*}

 \section{Discussion}
 \label{discussion}
In the present paper we have applied a one-zone leptohadronic model to six BL~Lacs, which
have been recently selected as the probable astrophysical counterparts (PR14) of five of the 
IceCube neutrino events \citep{aartsen14}, and calculated the neutrino flux in each case.
In contrast to other studies, where the neutrino emission is calculated under the assumption
of a generic blazar SED, here we obtained the neutrino signal from  BL~Lacs by fitting
their individual SEDs. 
Thus, for each source, the photon and particle distributions used to
derive the neutrino emission have to be determined 
individually, so as to give the observed SEDs.

% Our method suggests that the photon and particle distributions used
% to derive the neutrino emission are determined specifically for each source.

For the SED modelling, we have adopted one of the possible leptohadronic model variants that allows
us to directly associate the $\gamma$-ray  blazar emission with a high-energy ($2-20$~PeV)
neutrino signal. In particular, the low-energy hump of the SED is explained by synchrotron radiation of primary
relativistic electrons, whereas the observed high-energy (GeV-TeV) emission is the combined result
of synchrotron self-Compton emission (from primary electrons) and of 
synchrotron radiation from secondary pairs.
These are produced in a variety of ways, such as in Bethe--Heitler and $\gamma \gamma$ pair production, and through 
the decays of charged pions, which themselves are the  by-product of $\pg$ interactions 
of co-accelerated protons with the internally produced synchrotron photons. 
Secondary pairs produced in different processes have,  in general, different
energy distributions. Thus, in our framework, it is mainly the synchrotron emission 
of $\pg$ pairs that falls in the $\gamma$-ray regime, and in combination with the SSC radiation
is used to explain the observed blazar emission in $\gamma$-rays.

BL~Lacs are known for their variability across the electromagnetic spectrum and, because of this, 
we  used SEDs that are comprised of nearly simultaneous MW observations. For those sources that were
targets of recent MW observing campaigns we have adopted the data from the respective publications.
 In any other case,
we have compiled the SEDs using the SED builder tool of the ASI Science Data Centre (ASDC).
Only for one source in the sample (1H~1914-194), which had the poorest simultaneous MW coverage, had we to
construct an SED using non-simultaneous data. 
We used the numerical code described in DMPR12 in order to calculate steady-state photon spectra
that were later applied to the MW observations. 
Because of the multi-parameter nature of the problem (11-13 free parameters) and of the long numerical simulation time
required for every parameter set, we have not performed a blind search of the available parameter space.
To obtain a first estimate of the required parameter set, 
we used instead the analytical expressions presented in PM15.
We then performed a series of numerical simulations
using parameter values lying close to the initial parameter set,
until a reasonably good fit to each SED was obtained.

Although the model-derived neutrino emission
from the six BL~Lacs that were selected as counterparts of the IceCube events
was  the main focus of this paper, our study highlighted also
the flexibility of the leptohadronic model and its success in describing the blazar SEDs. 
This is particularly important, if one takes into account the following:
(i) all sources in the sample were fitted using
reasonable parameter values; (ii) these BL~Lacs were not  {\sl a priori} selected for modelling purposes;
and (iii) besides the fact that all sources in the sample were HBLs, 
they differed in many aspects, e.g. in redshift, in $\gamma$-ray luminosity and spectra.

The only source that may be excluded, at the present time, from  being the astrophysical counterpart of the detected neutrino (event 19)
is \rxs. \ In all other cases, the model-predicted neutrino flux is
below the 1~$\sigma$ but still within the 3~$\sigma$ error bars of the respective IceCube neutrino events.
In particular, for blazars Mrk~421 and 1H~1914-194 
the model-derived neutrino fluxes at the same energy with the respective neutrino events (ID 9 and 22, respectively)
were found to be very close to the low 1~$\sigma$ error bars (Fig.~\ref{neutrinos}).  Although
our results cannot be considered a proof of a firm association between these BL~Lacs and the respective neutrinos, 
they can be verified or disputed in the near future, as IceCube collects more data and the observed fluxes become  
more constraining for the model.

Among the blazars of the sample, Mrk~421 is a special case. Its neutrino emission was calculated 
in a previous study (DPM14), at a time when there was no hint of a possible association between Mrk~421 and the IceCube events.
DPM14  found that the neutrino flux from Mrk~421 was close
to the current IceCube (IC-40) sensitivity limit for this source (left panel in Fig.~\ref{mrk421}). Of particular interest are
the following:
\begin{itemize}
 \item the prediction of DPM14 was found {\sl a posteriori} to be close to the observed flux related to neutrino ID 9;
 \item  the same parameter set, besides the Doppler factor,  used in DPM14  to fit the optical/X-ray/TeV data of 2001, has been successfully 
  (right panel in Fig.~\ref{mrk421}) applied to the more complete dataset of the 2009 campaign \citep{abdo10}.
 \end{itemize}

Apart from a direct comparison of the model and observed neutrino fluxes, we have shown that we can establish
a direct connection between the observed $\gamma$-ray emission and the putative neutrino emission 
from a BL~Lac source. We have quantified this relation by introducing 
the ratio of the total neutrino to the $0.01-1$ TeV luminosity ($\ygn$).
In our model, for a given observed $\gamma$-ray luminosity that 
is purely explained by the synchrotron emission of $\pg$ pairs, i.e.
when the SSC contribution is negligible, there is an upper limit to the neutrino
luminosity. It can be shown that the limit is $\sim 3 L_{\gamma, \rm TeV}$ (see PM15).
If, however, the SSC component is dominant in the $\gamma$-ray regime, we expect $\ygn \rightarrow \epsilon \ll 1$; in this case,
our model simplifies into a leptonic one. Thus, the spectrum of possible values is
 $[\epsilon \ll 1, 3]$ and corresponds to different contributions of the $\pg$ and SSC components to the $\gamma$-ray emission.
 
For the six blazars we have found ratios in the range $0.1 \lesssim \ygn\lesssim 2$, with a mean value 
$\bar{Y}_{\nu \gamma} \simeq 0.8$. We have identified two sub-groups
in the sample: blazars Mrk~421, H~2356-309 and 1H~1914-194 with $\ygn \ge \bar{Y}_{\nu \gamma}$ and the rest
with ratios below the average value (Table~\ref{tab-0}).
Despite the small size of our sample, we have investigated the dependence of this ratio on observables of blazar emission.
No  hints of correlation with the photon index in the Fermi/LAT energy range were found (Fig.~\ref{Y_phindex}, top panel). 
However, we showed that there is a negative trend between $L_{\gamma, \rm TeV}$ and $\ygn$, i.e. more $\gamma$-ray luminous BL~Lacs
have a smaller photohadronic contribution  to their $\gamma$-ray spectra (bottom panel in Fig.~\ref{Y_phindex}). 
Although this could be interpreted as a lower injection luminosity of relativistic protons, we showed
that this is not the reason (see Fig.~\ref{LpLe} and Table~\ref{tab-2}). Instead, 
the separation of sources according to $\ygn$ can be  understood
in terms of the magnetic compactness. In Fig.~\ref{lB} we demonstrated the transition 
from SSC to $\pg$ dominated $\gamma$-ray emission as the
magnetic compactness gradually increases, which also translates to an increase of $\ygn$.
Application of the model to a larger sample of BL~Lacs is, however, necessary for 
a better understanding of the aforementioned trends.

Our analysis showed that the shape of the neutrino spectrum does not
strongly depend on the $\gamma$-ray luminosity. In all cases,
the neutrino spectra are described as $dN_{\nu}/dE_{\nu} \propto E_{\nu}^{\alpha -2}$ 
with $\alpha \simeq 1.1-1.3$ (Fig.~\ref{neutrinos}). The peak energy of the neutrino spectrum, in the present context,
depends only on one model parameter, namely the Doppler factor, and two observables, i.e. 
the peak frequency ($\nu_{\rm s}$) of the low-energy hump of the SED and the redshift of the source $z$. The dependence on the latter is not
strong.
Since all the sources of the sample are HBLs and we have obtained fits with similar values of $\delta$, we should
not expect large differences in the peak neutrino energy. We have verified this numerically, as demonstrated
in Fig.~\ref{neutrinos}. In addition to the similarity in the spectral shape, we have
connected the expected neutrino luminosity from a BL~Lac with
its $\gamma$-ray luminosity in the 0.01-1 TeV energy range. Thus, one could go one step further and  
calculate the diffuse neutrino flux from BL~Lacs, in order to see how it compares
with the current IceCube detections and upper limits in the sub-EeV regime.
Such a calculation requires an extrapolation of our
findings to the wider class of BL~Lacs, including low-synchrotron peaked blazars (LBLs), and the use
of detailed simulations of the blazar population of the type done by \cite{padovanigiommi15}. 
%PLEASE ADD THE RELEVANT REF. http://cdsads.u-strasbg.fr/abs/2015MNRAS.446L..41P
%\maria{Paolo could you add something here?}
This will be the focus of a future study.

Similarly to other leptohadronic models, such as the proton synchrotron model (e.g. \citealt{boettcherreimer13}),
the derived jet power is high, ranging from $10^{48}$~erg/s up to a few $10^{49}$~erg/s, which 
implies a very low radiative efficiency. 
Moreover, the emitting region is far from equipartition, with the proton energy density being the dominant one -- 
see Tables~\ref{tab-2} in this work and
\cite{boettcherreimer13}. The demanding energetics and the large departure from
equipartition constitute the main point for criticism of our model, and of leptohadronic models in general.
However, there are some important differences
compared to the proton synchrotron model, which make this leptohadronic variant (LH$\pi$) more flexible. 
On the one hand, strong magnetic fields ($\gg 10$~G) are not necessary, while, on the other hand,
proton acceleration to ultra-high energies ($\gamma_{\rm p} \sim 10^9-10^{10}$) is not required. On the contrary, the  upper cutoff of 
the proton distribution should be just a few times higher than $\sim 3.5 \times 10^7 (1+z)^{-1} \delta_1 \nu_{\rm s, 16}^{-1}$ 
(see equation~(\ref{eq2})). In terms of the available parameter space of leptohadronic models, 
the LH$\pi$ variant occupies a sub-region, which corresponds, roughly speaking, to: low-to-moderate magnetic field  strengths, 
moderate maximum proton energies, sub-pc sizes of the emission region ($\sim 10^{15}-10^{16}$~cm) and high jet powers.

% We investigated a leptohadronic variant where the blazar $\gamma$-ray emission has 
% a mixed origin, namely both photohadronic and SSC processes contribute to it. This 
% has been also recently discussed in 
% \cite{cerrutietal14}, although in the context of extreme BL~Lacs \citep{costamanteetal01}, yet without
% discussing the accompanying neutrino emission. However

A detailed calculation of the neutrino emission from blazars following the so-called ``blazar sequence'' \citep{ghisellinietal98} has been 
presented in two recent studies by \cite{murase14} and \cite{dermermurase14}. Our analysis showed, however, that
relatively high neutrino luminosities (of the order of $\sim 10^{45}$~erg/s) can be obtained from classical BL~Lacs, with
Mrk~421 being the most notable case. We note also that the $\pg$ production efficiency is low in all six blazars
(see Table~\ref{tab-0}), and is in agreement
with the expected values from BL~Lacs (see equation~(20) in \citealt{murase14} and equation~(28) in PM15). 
The question is then whether and how those results are compatible with each other. Figure~9 in \cite{murase14} shows
that the typical neutrino luminosity of individual BL~Lacs with no contamination by external photon fields  (two lower blue curves)
are $\sim 10^{42}$~erg/s, which is approximately 3 orders of magnitude below our typical values. The reason behind 
this discrepancy is the luminosity that is being injected into relativistic protons. In our analysis, we systematically 
used 2 to 3 orders of magnitude higher proton injection luminosities than \cite{murase14}. This is mainly due to
the fact that we try to model the observed  blazar $\gamma$-ray emission in terms of photohadronic processes, 
whereas in \cite{murase14} and \cite{dermermurase14} it is not clear what the fate of the photons produced via photohadronic processes is, and why
they do not appear in the observed SED. 
In any case, we compensate for the low number density of internal radiation by pushing the proton injection luminosity to higher values.

Another way to go around the problem of low photon number density in BL~Lacs, which is related to low neutrino production efficiency,
has been also recently presented in \cite{tavecchioetal14}. Their approach is different than ours, though. In their model, BL~Lacs can still be efficient neutrino
emitters if  the number density of photons, which are the targets for $\pg$ interactions, is enhanced. 
In order to achieve this, \cite{tavecchioetal14} assume
a structured BL~Lac jet with a fast spine and slow layer 
\citep{ghisellinietal05, tavecchioghisellini08}, where the target field for $\pg$ interactions
is the radiation field from the layer that appears boosted in the rest frame of the spine. 

The recent correlation analysis of IceCube neutrino events and $\gamma$-ray sources by PR14 led to a somewhat
unexpected result: instead of the more powerful FSRQs, it was  BL~Lacs that 
were among the most plausible astrophysical counterparts of 9 IceCube neutrinos. 
In general, FSRQs are expected to be efficient PeV neutrino emitters, because of their higher $\gamma$-ray
luminosity, but more importantly, because of the dense external photon fields (broad line region and/or dusty torus).
The absence of FSRQs from the list of probable astrophysical counterparts may be
related to the ``energetic criterion'' applied by PR14. This was totally empirical and observationally 
driven, since it assumed only a
tight photon -- neutrino connection; any other more sophisticated choice of a diagnostic would be model-dependent. 
Thus, sources where a simple extrapolation of
the $\gamma$-ray flux to the PeV energies lied much below the observed neutrino flux (taking into account
the uncertainty in the flux measurement) were discarded. Relaxing this constraint would certainly
increase the number of candidates, which might include FSRQs, but these would
then depend on the specific model assumed.

In principle, the leptohadronic variant we presented here can also be applicable to 
FSRQs, with the appropriate choice of parameters. Because of the higher $\gamma$-ray luminosities and lower peak energies of FSRQs with respect 
to HBLs, a straightforward way to apply our model 
to FSRQs would be to increase the proton injection luminosity, while lowering the maximum proton energy. Taking into account the presence of additional 
photons originating from the broad line region and/or the torus, the model-predicted (PeV) neutrino luminosity would be much higher than the one
derived for the HBLs in our sample and possibly  close to, or at, the limit of flux detected by IceCube.
% One the one hand, this scenario is promising for the neutrino production, while, on the other hand, is challenging
% for the ``energetic criterion'' applied by PR14 while searching for the most plausible astrophysical counterparts of the IceCube neutrinos.
The application of the model to the SEDs of FSRQs, however, may not be as simple as described above. Higher $L'_{\rm p, inj}$ 
leads not only to increased emission from $\pg$ interactions but also from Bethe-Heitler interactions. Given its spectral shape
and the typical energy range that emerges (see Figs.~\ref{mrk421}-\ref{rxs-comp}), it would be difficult to reconcile its emission
with the flat X-ray spectra of FSRQs. 
However, more concrete answers can be given only with a detailed fitting of these type of sources.

 \section{Summary}
 \label{summary}
Following the identification of eight BL Lac objects as likely
sources of IceCube neutrinos by PR14, we set out to deduce the neutrino
emission of the ones whose redshift is known by applying a leptohadronic 
model, using a variant of the model first discussed in \cite{petromast12}, albeit in a
more general context. We used data 
from near simultaneous MW observations for all sources (as far as possible), and fitted the 
resulting SEDs in each individual case with steady-state photon spectra, which
were obtained with the DMPR12 numerical code.
Through the fitting procedure we determined the distributions of injected electrons and 
protons in what we assumed were spherical volumes of certain radii and magnetic field strengths, 
moving towards us with certain Lorentz factors; all the above being treated as free parameters.

Of the six BL Lacs with known redshift that we studied, one of them, Mrk~421, 
had been considered for its potential as a neutrino source in the past, by DPM14; 
the present study verifies that the neutrino event associated with it (ID 9) falls within 
its predicted flux. A good match between modeled and observed neutrino (ID 22) flux was also found for 
1H~1914-194. 1~RXS J054357.3-553206, on the other hand, was the only BL lac to be confidently excluded as a 
source of its associated neutrino (ID 19), with the remaining three BL Lacs displaying some discrepancy between modeled 
and observed fluxes, but not enough to rule them out as sources at this stage.  Depending on the shape of their 
SEDs, we found that some cases favour fits dominated by SSC while others favour fits dominated 
by photohadronic interactions, 
with the former having a lower ratio of neutrino to $0.01-1$ TeV $\gamma$-ray luminosities than the latter. 
This ratio, $Y_{\nu\gamma}$, which could prove important for calculations of the diffuse neutrino
emission in the future, was then evaluated for its dependence on observable and model 
parameters. Ultimately, the results of our model should be confirmed or disproven in the near 
future, with the accumulation and data analysis of additional IceCube observations. Whatever the 
outcome, it will be important for constraining the available 
parameter space for leptohadronic models of blazar emission. 
This illustrates the importance of multi-messenger high-energy astronomy.

\section*{Acknowledgments}
MP was supported by NASA 
through Einstein Postdoctoral 
Fellowship grant number PF3~140113 awarded by the Chandra X-ray 
Center, which is operated by the Smithsonian Astrophysical Observatory
for NASA under contract NAS8-03060. 
ER acknowledges the support of the Deutsche Forschungsgemeinschaft (DFG)
\bibliographystyle{mn2e} % style mn2e.bst
\bibliography{modelling.bib}

\end{document}